\documentclass[prd,twocolumn,floatfix,amsmath,nofootinbib,amssymb,floatfix]{revtex4}
\usepackage{graphicx,color,dcolumn,booktabs,bm}
\usepackage{longtable,lscape}
\usepackage{pdfpages}
\usepackage{txfonts}
\usepackage{overpic}
\usepackage{amssymb}
\usepackage{makecell}
\usepackage{indentfirst}
\usepackage{feynmf}   %{feynmp}
\usepackage{slashed}  %for Feynman symbols
\usepackage{cases}
\usepackage{color}
\usepackage{multirow}
\usepackage{threeparttable}
\usepackage{epstopdf}
\usepackage{enumerate}
\usepackage{subfigure}
\usepackage{diagbox}
\usepackage{graphicx,color,dcolumn,booktabs,bm}
\usepackage{mathrsfs}
\usepackage{cancel}
\usepackage{float}
\usepackage[colorlinks,
            citecolor=blue,
            anchorcolor=red,
            menucolor=red,
            linkcolor=red,
            filecolor=red,
            runcolor=red,
            urlcolor=blue,
            frenchlinks=red]{hyperref}

\begin{document}
\title{Mass spectra of singly heavy baryons in the relativized quark model with heavy-quark dominance}
\author{Zhen-Yu Li$^{1,4}$}
\email{zhenyvli@163.com }
\author{Guo-Liang Yu$^{2}$}
\email{yuguoliang2011@163.com }
\author{Zhi-Gang Wang$^{2}$ }
\email{zgwang@aliyun.com }
\author{Jian-Zhong Gu$^{3}$ }
\email{gujianzhong2000@aliyun.com }
\author{Hong-Tao Shen$^{4}$ }
\email{shenht@gxnu.edu.cn }

\affiliation{$^1$ School of Physics and Electronic Science, Guizhou Education University, Guiyang 550018,
China\\$^2$ Department of Mathematics and Physics, North China Electric Power University, Baoding 071003,
China\\$^3$ China Institute of Atomic Energy, Beijing 102413,China\\$^4$ Guangxi Key Lab of Nuclear Physics and Technology, Guangxi Normal University, Guilin 541006, China}
\date{\today }

\begin{abstract}
The rigorous calculation of the spin-orbit terms in the three-quark system is realized based on the Gaussian expansion method and the infinitesimally-shifted Gaussian basis functions in the framework of the relativized quark model, by ignoring the mixing between different excited states. Then, the complete mass spectra of the singly heavy baryons are obtained rigorously, under the mechanism of the heavy-quark dominance. On these bases, the systematical analyses are carried out for the reliability and predictive power of the model, the fine structure of the singly heavy baryon spectra, the assignments of the excited baryons, and some important topics about the heavy baryon spectroscopy such as the missing states, the `spin-orbit puzzle', the clustering effect, etc. The result confirms that under the heavy-quark dominance mechanism, the relativized quark model can describe the excitation spectra and the fine structures of the singly heavy baryons correctly and precisely.

\end{abstract}

\maketitle

\section{Introduction}\label{sec1}

The heavy baryon spectroscopy is crucial for gaining deeper insights into the strong interaction in
the non-perturbative regime of the Quantum Chromodynamics (QCD)~\cite{F101}. It has attracted considerable experimental and theoretical attentions. So far, a large number of singly heavy baryons have been observed in experiment~\cite{F201,F2021,P2940,F205,P2910,P2882,F206,F207,F208,F210,F209,F211,F212,F213,F4401,F4402,F4403,F203,F204,F214,LHCb25}, which provides an important support for related theoretical researches~\cite{F411,Chenhx23,Crede24}.

In the new Review of Particle Physics (RPP) by the Particle Data Group (PDG), more than 70 singly heavy baryons have been collected~\cite{F201}. These heavy baryons and their $J^{P}$ values are listed in Table~\ref{tta0001}, which shows that most of the ground states of the heavy baryons have been well established in experiment. But the $J^{P}$ values of many excited baryons have not been identified. Moreover, some of the excited baryons were observed experimentally in groups, and their mass values are very close to each other, such as $\{\Omega_{c}^{0}(3000)$, $\Omega_{c}^{0}(3050)$, $\Omega_{c}^{0}(3065)$, $\Omega_{c}^{0}(3090)$, $\Omega_{c}^{0}(3120)\}$, $\{\Xi_{c}(2923)^{0}$, $\Xi_{c}(2930)^{+,0}$, $\Xi_{c}(2970)^{+,0}\}$, and $\{\Omega_{b}(6316)^{-}$, $\Omega_{b}(6330)^{-}$, $\Omega_{b}(6340)^{-}$, $\Omega_{b}(6350)^{-}\}$. These close mass values in each group indicate a fine structure in their excitation spectra, which is, however, an unsolved problem in the current theory. In addition, lots of the excited heavy baryons as shown in Table~\ref{tta0001} have been observed in the last few years, due to the improvement of experimental accuracy by some collaborations such as the LHCb, the Belle, the CMS, etc. Very recently, a new charmed baryon $\Xi_{c}(2923)^{+}$ was firstly observed by the LHCb collaboration~\cite{LHCb25}. It is expected that more heavy baryons will be observed in the near future, and more fine structures are also expected to be discovered.

\begin{table*}[htbp]
\begin{ruledtabular}\caption{Observed singly heavy baryons and their $J^{P}$ values~\cite{F201}. The $\Sigma_{c}(2846)^{0}$ and the $\Xi_{c}(2923)^{+}$ are cited from Ref.~\cite{F209} and Ref.~\cite{LHCb25}, respectively. }
\label{tta0001}
\begin{tabular}{c c c c c c c c c c c c c c c c c c c c c c c}
Baryon  & $J^{P}$   &Baryon  & $J^{P}$   &Baryon  & $J^{P}$   &Baryon  & $J^{P}$   &Baryon  & $J^{P}$  &Baryon  & $J^{P}$ &Baryon  & $J^{P}$ &Baryon  & $J^{P}$ \\\hline
$\Lambda_{c}^{+}$ & $\frac{1}{2}^{+}$ &$\Sigma_{c}(2455)^{++}$ & $\frac{1}{2}^{+}$ &$\Xi_{c}^{+}$ & $\frac{1}{2}^{+}$ &$\Omega_{c}^{0}$ & $\frac{1}{2}^{+}$&$\Lambda_{b}^{0}$ & $\frac{1}{2}^{+}$&$\Sigma_{b}^{+}$ & $\frac{1}{2}^{+}$ &$\Xi_{b}^{0}$ & $\frac{1}{2}^{+}$&$\Omega_{b}^{-}$ & $\frac{1}{}^{+}$ \\

$\Lambda_{c}(2595)^{+}$ & $\frac{1}{2}^{-}$ &$\Sigma_{c}(2455)^{+}$ & $\frac{1}{2}^{+}$ &$\Xi_{c}^{0}$ & $\frac{1}{2}^{+}$ &$\Omega_{c}(2770)^{0}$ & $\frac{3}{2}^{+}$ &$\Lambda_{b}(5912)^{0}$ & $\frac{1}{2}^{-}$ &$\Sigma_{b}^{-}$ & $\frac{1}{2}^{+}$  &$\Xi_{b}^{-}$ & $\frac{1}{2}^{+}$&$\Omega_{b}(6316)^{-}$ & $?^{?}$\\
$\Lambda_{c}(2625)^{+}$ & $\frac{3}{2}^{-}$ &$\Sigma_{c}(2455)^{0}$  & $\frac{1}{2}^{+}$ &$\Xi_{c}^{'+}$ & $\frac{1}{2}^{+}$ &$\Omega_{c}(3000)^{0}$ & $?^{?}$ &$\Lambda_{b}(5920)^{0}$ & $\frac{3}{2}^{-}$ &$\Sigma_{b}^{*+}$ & $\frac{3}{2}^{+}$ &$\Xi_{b}(5935)^{-}$ & $\frac{1}{2}^{+}$&$\Omega_{b}(6330)^{-}$ & $?^{?}$\\
$\Lambda_{c}(2765)^{+}$ & $?^{?}$     &$\Sigma_{c}(2520)^{++}$ & $\frac{3}{2}^{+}$ &$\Xi_{c}^{'0}$ & $\frac{1}{2}^{+}$ &$\Omega_{c}(3050)^{0}$ & $?^{?}$ &$\Lambda_{b}(6070)^{0}$ & $\frac{1}{2}^{+}$&$\Sigma_{b}^{*-}$ & $\frac{3}{2}^{+}$ &$\Xi_{b}(5945)^{0}$ & $\frac{3}{2}^{+}$&$\Omega_{b}(6340)^{-}$ & $?^{?}$\\
$\Lambda_{c}(2860)^{+}$ & $\frac{3}{2}^{+}$   &$\Sigma_{c}(2520)^{+}$ & $\frac{3}{2}^{+}$  &$\Xi_{c}(2645)^{+}$ & $\frac{3}{2}^{+}$ &$\Omega_{c}(3065)^{0}$ & $?^{?}$ &$\Lambda_{b}(6146)^{0}$ & $\frac{3}{2}^{+}$&$\Sigma_{b}(6097)^{+}$ & $?^{?}$ &$\Xi_{b}(5955)^{-}$ & $\frac{3}{2}^{+}$&$\Omega_{b}(6350)^{-}$ & $?^{?}$\\
$\Lambda_{c}(2880)^{+}$ & $\frac{5}{2}^{+}$  &$\Sigma_{c}(2520)^{0}$ & $\frac{3}{2}^{+}$  &$\Xi_{c}(2645)^{0}$ & $\frac{3}{2}^{+}$  &$\Omega_{c}(3090)^{0}$ & $?^{?}$ &$\Lambda_{b}(6152)^{0}$ & $\frac{5}{2}^{+}$&$\Sigma_{b}(6097)^{-}$ & $?^{?}$ &$\Xi_{b}(6087)^{0}$ & $\frac{3}{2}^{-}$\\
$\Lambda_{c}(2910)^{+}$ & $?^{?}$   &$\Sigma_{c}(2800)^{++}$ & $?^{?}$   &$\Xi_{c}(2790)^{+}$ & $\frac{1}{2}^{-}$  &$\Omega_{c}(3120)^{0}$ & $?^{?}$ &&&&&$\Xi_{b}(6095)^{0}$ & $\frac{3}{2}^{-}$\\
$\Lambda_{c}(2940)^{+}$ & $\frac{3}{2}^{-}$    &$\Sigma_{c}(2800)^{+}$ & $?^{?}$  &$\Xi_{c}(2790)^{0}$ & $\frac{1}{2}^{-}$ &$\Omega_{c}(3185)^{0}$ & $?^{?}$ &&&&&$\Xi_{b}(6100)^{-}$ & $\frac{3}{2}^{-}$\\
                             &    &$\Sigma_{c}(2800)^{0}$ & $?^{?}$ &$\Xi_{c}(2815)^{+}$ & $\frac{3}{2}^{-}$&$\Omega_{c}(3327)^{0}$ & $?^{?}$ &&&&&$\Xi_{b}(6227)^{0}$ & $?^{?}$\\
                                                         &     &$\Sigma_{c}(2846)^{0}$ & $?^{?}$ &$\Xi_{c}(2815)^{0}$ & $\frac{3}{2}^{-}$&& &&&&&$\Xi_{b}(6227)^{-}$ & $?^{?}$\\
                                                              &    & &  &$\Xi_{c}(2882)^{0}$ & $?^{?}$&& &&&&&$\Xi_{b}(6327)^{0}$ & $?^{?}$\\
                                                              &     & &  &$\Xi_{c}(2923)^{+}$ & $?^{?}$&& &&&&&$\Xi_{b}(6333)^{0}$ & $?^{?}$\\
                                                               &    & & &$\Xi_{c}(2923)^{0}$ & $?^{?}$\\
                                                               &    & & &$\Xi_{c}(2930)^{+}$ & $?^{?}$\\
                                                               &     & &  &$\Xi_{c}(2930)^{0}$ & $?^{?}$\\
                                                                  &     & &  &$\Xi_{c}(2970)^{+}$ & $\frac{1}{2}^{+}$\\
                                                                    &    & &  &$\Xi_{c}(2970)^{0}$ & $\frac{1}{2}^{+}$\\
                                                                      &     & &  &$\Xi_{c}(3055)^{+}$ & $?^{?}$\\
                                                                        &     & &  &$\Xi_{c}(3080)^{+}$ & $?^{?}$\\
                                                                          &    & &  &$\Xi_{c}(3080)^{0}$ & $?^{?}$\\
                                                                          &    & &  &$\Xi_{c}(3120)^{+}$ & $?^{?}$
\end{tabular}
\end{ruledtabular}
\end{table*}

All these experimental progresses show that it is time to systematically analyze the data and delineate a reliable mass spectrum.
However, it is not a simple matter to give an accurate analysis of these observed heavy baryons theoretically, which has actually become a great challenge for various theoretical methods. As an indispensable tool for understanding of the multitude of observed baryons and their properties, the relativized quark model with QCD also faces the same challenge.

The relativized quark model was developed by Godfrey and Isgur in 1985~\cite{F401}, and has achieved great success in analyzing the meson spectra. The Hamiltonian of this model is based on a universal one-gluon-exchange-plus-linear-confinement potential motivated by QCD, which contains almost all possible forms of the main interaction between the two quarks. In 1986, Capstick and Isgur extended this model and insisted on using the method of studying light-quark baryons and systematically studied the mass spectra of both light and heavy baryons under a unified framework~\cite{F402}. Their study in the baryon spectroscopy produced a lasting effect~\cite{Fp003}. However, their study predicted more `missing' states of the heavy baryons, which is very similar to the case of the light-quark baryons.
Once more, in a similar manner to the light-quark baryons, there are two possible solutions to the problem for the heavy baryons summarized by Capstick and Roberts. The first one is that the dynamical degrees of freedom used in the model, namely the three valence quarks, are not physically realized. Instead, a baryon consists of a quark and a diquark, and the reduction of the number of internal degrees of freedom leads to a more sparsely populated spectrum. The second possible solution is that the missing states couple weakly to the formation channels used to investigate the states, and so give very small contributions to the scattering cross sections~\cite{F306}.

Later, the heavy quark symmetry~\cite{Isgur1991}, the heavy quark limit~\cite{F304} and the heavy quark effective theory~\cite{F303,F403} were put forward one after another,  and revealed some important structure properties of the heavy baryons, which laid the foundation for the solution of the above problem. According to the first possible solution, Ebert, Faustov and Galkin analyzed the spectra of the singly heavy baryons in the heavy quark-light diquark picture~\cite{F405}, and predicted significantly fewer states than those of Ref.~\cite{F402} mentioned above, which has two important implications. One is that the total orbital angular momentum $L$ can be approximatively regarded as a good quantum number of a baryon state, even though it is not true strictly in a relativistic theory. In practice, as an approximative good quantum number, $L$ has been widely used in researches~\cite{F315,f17a0,f17p13,f17,f17p10,f17p8,f17p14,f17p7,f17p11}. An other is that the concept of `the clustering effect' is officially applied in study, which means there might exist the cluster in the singly heavy baryon, if this solution is correct.
However, the reliability of the first solution has yet to be tested further. `It is telling that this simple diagnostic is difficult to apply since so little is known of the excited baryon spectrum'~\cite{F101}.

Inspired by the above related theoretical works, we studied the spectra of the singly and doubly heavy baryons systematically in the framework of the relativized quark model~\cite{F502,F503,F504,F505,F506}. The used method adopted the respective advantages of the above two possible solutions. We considered $L$ to be an approximative good quantum number, assumed the stable (or physically realized) quantum states for the excited heavy baryons should live in the lower orbital excitation mode, and further ignored the mixing between different excited states. The results showed that most of the experimental data can be well described with a uniform set of parameters for the heavy baryons. We analyzed the orbital excitation of the heavy baryons carefully and proposed the heavy-quark dominance (HQD) mechanism, which may solve the problem of the `missing' states in a natural way, and determine the overall structure of the excitation spectra for the singly and doubly heavy baryons~\cite{F501}.

For describing the fine structure of the observed excited baryons, we improved the calculation of the spin-orbit interactions by considering the contribution from the light-quark cluster in a quasi-two-body spin-orbit interaction, which enhances the energy level splitting of the orbital excitation significantly and presents a reasonable fine structure~\cite{F507}.
The analysis of the fine structure confirms that the contribution of the spin-orbit interaction from the orbital angular momentum $\textbf{\emph{l}}_{\lambda}$ is not negligible.

The predicted singly heavy baryon spectra in our works match well with the current data. But, it is still unsatisfactory because the approximate formulas were used for describing the contributions of the spin-orbit interaction to the fine structures~\cite{F507}, as a result, one cannot judge the deviation from the real results. This reduces the reliability of the calculation and the predictive power. So, it is necessary to analyze the fine structure by using the rigorous calculation. However, the rigorous calculation is a common tough problem in the three-body systems.
Because the Hamiltonian of the relativized quark model is based on the two-body interaction, one will encounter some technical difficulties in the rigorous calculation, when the model is extended from the mesons to the baryons. This is indeed the biggest obstacle that this model has encountered in studying the three-quark systems. If the rigorous calculation is implemented, some important problems of this model appearing in the heavy baryon spectroscopy might be solved, such as the missing states~\cite{F306}, the `spin-orbit puzzle'~\cite{Fp001,Fp002}, the clustering effect in a heavy baryon, etc. And a more important question could also be answered, i.e., whether and how the relativized quark model can correctly describe the heavy baryon spectroscopy.

In this work, we will try to perform the rigorous calculation of the heavy baryon spectra in the relativized quark model with the HQD mechanism, by using the Gaussian
expansion method (GEM) and the infinitesimally-shifted Gaussian (ISG) basis functions~\cite{F6021,F602}, so as to obtain a complete mass spectrum of the singly heavy baryons, answer the questions mentioned above and provide a reliable analysis for the relative researches.

The remainder of this paper is organized as follows. In
Sec.~\ref{sec2}, the theoretical methods used in this work are introduced, including the Hamiltonian of the relativized quark model, the wave functions and the Jacobi coordinates, and the evaluations of the matrix elements, including the rigorous calculation of the spin-orbit terms. The structure properties of the singly heavy baryon spectra, the comparison between the calculated excitation spectra and the experimental data, and the reliability of the model are analyzed in Sec.~\ref{sec3}. And Sec.~\ref{sec4} is reserved for the conclusions.

\section{Theoretical methods used in this work}\label{sec2}

\subsection{Hamiltonian of the relativized quark model}\label{sec2.1}
In the relativized quark model, the Hamiltonian for a three-quark system is based on the two-body interactions,
\begin{eqnarray}\label{e1}
\notag
H &&=H_{0}+\tilde{H}^{conf}+\tilde{H}^{hyp}+\tilde{H}^{so}\\
&&=\sum_{i=1}^{3}\sqrt{p_{i}^{2}+m_{i}^{2}}+\sum _{i<j}(\tilde{H}^{conf}_{ij}+\tilde{H}^{hyp}_{ij}+\tilde{H}^{so}_{ij}),
\end{eqnarray}
where the interaction terms $\tilde{H}^{conf}_{ij}$, $\tilde{H}^{hyp}_{ij}$ and $\tilde{H}^{so}_{ij}$ are the confinement, hyperfine and spin-orbit interactions, respectively.
The confinement term $\tilde{H}^{conf}_{ij}$ includes a modified one-gluon-exchange potential $G'_{ij}(r)$ and a smeared linear confinement potential $\tilde{S}_{ij}(r)$. The hyperfine interaction $\tilde{H}^{hyp}_{ij}$ consists of the tensor term $\tilde{H}^{tensor}_{ij}$ and the contact term $\tilde{H}^{c}_{ij}$. And the spin-orbit interaction $\tilde{H}^{so}_{ij}$ can be divided into the color-magnetic term $\tilde{H}^{so(v)}_{ij}$ and the Thomas-precession term $\tilde{H}^{so(s)}_{ij}$.
Their forms are described in detail below.
 \begin{eqnarray}
 \begin{aligned}
&\tilde{H}^{conf}_{ij}=G'_{ij}(r)+\tilde{S}_{ij}(r), \\
&\tilde{H}^{hyp}_{ij}=\tilde{H}^{tensor}_{ij}+\tilde{H}^{c}_{ij},\\
&\tilde{H}^{so}_{ij}=\tilde{H}^{so(v)}_{ij}+\tilde{H}^{so(s)}_{ij},
\end{aligned}
\end{eqnarray}
with
 \begin{eqnarray}
 &\tilde{H}^{tensor}_{ij}=-\frac{\textbf{s}_{i}\cdot\textbf{r}_{ij}\textbf{s}_{j}\cdot\textbf{r}_{ij}/r^{2}_{ij}-\frac{1}{3}\textbf{s}_{i}\cdot\textbf{s}_{j}}{m_{i}m_{j}}
\times(\frac{\partial^{2}}{\partial{r^{2}_{ij}}}-\frac{1}{r_{ij}}\frac{\partial}{\partial{r_{ij}}})\tilde{G}^{t}_{ij}, \\
&\tilde{H}^{c}_{ij}=\frac{2\textbf{s}_{i}\cdot\textbf{s}_{j}}{3m_{i}m_{j}}\nabla^{2}\tilde{G}^{c}_{ij},\\
\label{e5}
&\tilde{H}^{so(v)}_{ij}=\frac{\textbf{s}_{i}\cdot\textbf{L}_{(ij)i}}{2m^{2}_{i}r_{ij}}\frac{\partial\tilde{G}^{so(v)}_{ii}}{\partial{r_{ij}}}+
\frac{\textbf{s}_{j}\cdot\textbf{L}_{(ij)j}}{2m^{2}_{j}r_{ij}}\frac{\partial\tilde{G}^{so(v)}_{jj}}{\partial{r_{ij}}}+
\frac{(\textbf{s}_{i}\cdot\textbf{L}_{(ij)j}+\textbf{s}_{j}\cdot\textbf{L}_{(ij)i})}{m_{i}m_{j}r_{ij}}\frac{\partial\tilde{G}^{so(v)}_{ij}}{\partial{r_{ij}}}, \\
\label{e6}
&\tilde{H}^{so(s)}_{ij}=-\frac{\textbf{s}_{i}\cdot\textbf{L}_{(ij)i}}{2m^{2}_{i}r_{ij}}\frac{\partial\tilde{S}^{so(s)}_{ii}}{\partial{r_{ij}}}-
\frac{\textbf{s}_{j}\cdot\textbf{L}_{(ij)j}}{2m^{2}_{j}r_{ij}}\frac{\partial\tilde{S}^{so(s)}_{jj}}{\partial{r_{ij}}}.
\end{eqnarray}
Here, the following conventions are used, i.e., $\textbf{L}_{(ij)i}=\mathbf{r}_{ij}\times\mathbf{p}_{i}$ and $\textbf{L}_{(ij)j}=-\mathbf{r}_{ij}\times\mathbf{p}_{j}$. In the formulas above, $G'_{ij}$, $\tilde{G}^{t}_{ij}$, $\tilde{G}^{c}_{ij}$, $\tilde{G}^{so(v)}_{ij}$ and $\tilde{S}^{so(s)}_{ii}$ should be modified with the momentum-dependent factors as follows,
\begin{eqnarray}
\begin{aligned}\label{e7}
&G'_{ij}=(1+\frac{p^{2}_{ij}}{E_{i}E_{j}})^{\frac{1}{2}}\tilde{G}_{ij}(r_{ij})(1+\frac{p^{2}_{ij}}{E_{i}E_{j}})^{\frac{1}{2}}, \\
&\tilde{G}^{t}_{ij}=(\frac{m_{i}m_{j}}{E_{i}E_{j}})^{\frac{1}{2}+\epsilon_{t}}\tilde{G}_{ij}(r_{ij})(\frac{m_{i}m_{j}}{E_{i}E_{j}})^{\frac{1}{2}+\epsilon_{t}},\\
&\tilde{G}^{c}_{ij}=(\frac{m_{i}m_{j}}{E_{i}E_{j}})^{\frac{1}{2}+\epsilon_{c}}\tilde{G}_{ij}(r_{ij})(\frac{m_{i}m_{j}}{E_{i}E_{j}})^{\frac{1}{2}+\epsilon_{c}},\\
&\tilde{G}^{so(v)}_{ij}=(\frac{m_{i}m_{j}}{E_{i}E_{j}})^{\frac{1}{2}+\epsilon_{so(v)}}\tilde{G}_{ij}(r_{ij})(\frac{m_{i}m_{j}}{E_{i}E_{j}})^{\frac{1}{2}+\epsilon_{so(v)}}, \\
&\tilde{S}^{so(s)}_{ii}=(\frac{m_{i}m_{i}}{E_{i}E_{i}})^{\frac{1}{2}+\epsilon_{so(s)}}\tilde{S}_{ij}(r_{ij})(\frac{m_{i}m_{i}}{E_{i}E_{i}})^{\frac{1}{2}+\epsilon_{so(s)}},
\end{aligned}
\end{eqnarray}
where $E_{i}=\sqrt{m^{2}_{i}+p^{2}_{ij}}$ is the relativistic kinetic energy, and $p_{ij}$ is the momentum magnitude of either of the
quarks in the center-of-mass frame of the $ij$ quark subsystem~\cite{F402,F601}.

$\tilde{G}_{ij}(r_{ij})$ and $\tilde{S}_{ij}(r_{ij})$ are obtained by the smearing transformations of the one-gluon exchange potential $G(r)=-\frac{4\alpha_{s}(r)}{3r}$ and
linear confinement potential $S(r)=\tilde{b}r+\tilde{c}$, respectively,
\begin{eqnarray}\label{e8}
&\tilde{G}_{ij}(r_{ij})=\textbf{F}_{i}\cdot\textbf{F}_{j} \sum^{3}_{k=1}\frac{2\alpha_{k}}{\sqrt{\pi}r_{ij}}\int^{\tau_{kij}r_{ij}}_{0}e^{-x^{2}}\mathrm{d}x,
\end{eqnarray}
\begin{eqnarray}\label{e9}
\notag
\tilde{S}_{ij}(r_{ij}) &&=-\frac{3}{4}\textbf{F}_{i}\cdot\textbf{F}_{j} \{\tilde{b}r_{ij}[\frac{e^{-\sigma^{2}_{ij}r^{2}_{ij}}}{\sqrt{\pi}\sigma_{ij} r_{ij}}\\
&&+(1+\frac{1}{2\sigma^{2}_{ij}r^{2}_{ij}})\frac{2}{\sqrt{\pi}}\int^{\sigma_{ij}r_{ij}}_{0}e^{-x^{2}}\mathrm{d}x]+\tilde{c} \},
\end{eqnarray}
with
\begin{eqnarray}
\begin{aligned}
&\tau_{kij}=\frac{1}{\sqrt{\frac{1}{\sigma^{2}_{ij}}+\frac{1}{\gamma^{2}_{k}}}}, \\
&\sigma_{ij}=\sqrt{s^{2}_{0}(\frac{2m_{i}m_{j}}{m_{i}+m_{j}})^{2}+\sigma^{2}_{0}[\frac{1}{2}(\frac{4m_{i}m_{j}}{(m_{i}+m_{j})^{2}})^{4}+\frac{1}{2}]}.
\end{aligned}
\end{eqnarray}
Here $\alpha_{k}$ and $\gamma_{k}$ are constants. $\textbf{F}_{i}\cdot\textbf{F}_{j}$ stands for the inner product of the color matrices of quarks $i$ and $j$. For the baryon, $\langle\textbf{F}_{i}\cdot\textbf{F}_{j}\rangle=-2/3$. All of the parameters in these formulas are completely consistent with those in our previous works~\cite{F502,F503}. Their values are listed in Table~\ref{tta001}.

\begin{table*}[htbp]
\begin{ruledtabular}\caption{Parameters of the relativized quark model in this work. Their values are the same as those in Ref.~\cite{F401}, apart from $\tilde{b}$ and $\tilde{c}$~\cite{F502}. }
\label{tta001}
\begin{tabular}{c c c c c c c c c c c}
$m_{u}$/$m_{d}$(GeV)& $m_{s}$(GeV)   & $m_{c}$(GeV) & $m_{b}$(GeV) & $\gamma_{1}$(GeV) & $\gamma_{2}$(GeV) & $\gamma_{3}$(GeV)  & $\tilde{b}$(GeV$^{2}$) & $\tilde{c}$(GeV) \\
0.22 & 0.419  & 1.628 & 4.977& $1/2$ & $\sqrt{10}/2$ & $\sqrt{1000}/2$  & \textbf{0.14} & \textbf{-0.198}   \\\hline
$\epsilon_{c}$ & $\epsilon_{t}$   & $\epsilon_{SO(v)}$ & $\epsilon_{SO(s)}$  & $\alpha_{1}$ & $\alpha_{2}$ & $\alpha_{3}$  & $\sigma_{0}$(GeV)& $\tilde{s}$   \\
-0.168 & 0.025  & -0.035 & 0.055  & 0.25 & 0.15 & 0.20 & 1.8 & 1.55    \\
\end{tabular}
\end{ruledtabular}
\end{table*}

\subsection{Wave functions and Jacobi coordinates}\label{sec2.2}

For a singly heavy baryon system, the heavy-quark is decoupled from the two light-quarks in the heavy quark limit. With the requirement of the flavor $SU(3)$ subgroups for the light-quark pair, the singly heavy baryons belong to either a sextet ($\mathbf{6}_{F}$) of the flavor symmetric states,
 \begin{eqnarray}
\begin{aligned}
&\Sigma_{Q}=(uu)Q,~\frac{1}{\sqrt{2}}(ud+du)Q,~(dd)Q, \\
&\Xi_{Q}^{'}=\frac{1}{\sqrt{2}}(us+su)Q,~\frac{1}{\sqrt{2}}(ds+sd)Q,\\
&\Omega_{Q}=(ss)Q,
\end{aligned}
\end{eqnarray}
or an anti-triplet ($\mathbf{\bar{3}}_{F}$) of the flavor antisymmetric states~\cite{F403},
\begin{eqnarray}
\begin{aligned}
&\Lambda_{Q}=~\frac{1}{\sqrt{2}}(ud-du)Q, \\
&\Xi_{Q}=\frac{1}{\sqrt{2}}(us-su)Q,~\frac{1}{\sqrt{2}}(ds-sd)Q.
\end{aligned}
\end{eqnarray}
Here $u$, $d$ and $s$ denote up, down and strange quarks, respectively. $Q$ denotes charm ($c$) quark or bottom ($b$) quark.

\begin{figure}[htbp]
\centering
\includegraphics[width=8.5cm]{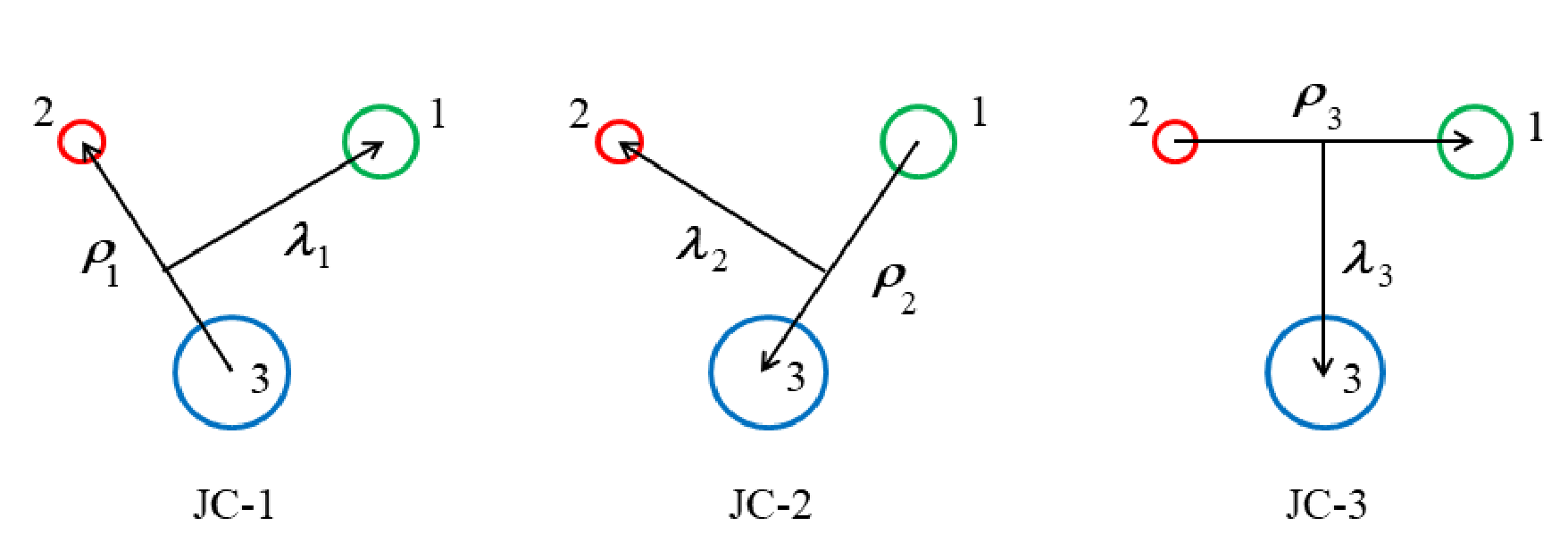}
\caption{There are 3 channels of the Jacobi coordinates for a three-quark system, labeled with $\{\boldsymbol\rho_{k}$, $\boldsymbol\lambda_{k}\}$ ($k$=1, 2, 3). The channel 3 (JC-3) is selected for defining the wave function of a singly heavy baryon state. All the quarks are numbered for ease of use in calculations, and the 3rd quark refers specifically to the heavy quark.}
\label{f1}
\end{figure}

For describing the internal orbital motion of the singly heavy baryon, we select the specific Jacobi coordinates (named JC-3 for short) as shown in Fig.~\ref{f1}, which is consistent with the above reservation about the flavor wave function naturally.
In this work, the Jacobi coordinates are defined as
\begin{eqnarray}
\begin{aligned}
&\boldsymbol\rho_{i}=\textbf{r}_{jk}=\textbf{r}_{j}-\textbf{r}_{k}, \\
&\boldsymbol\lambda_{i}=\textbf{r}_{i}-\frac{m_{j}\textbf{r}_{j}+m_{k}\textbf{r}_{k}}{m_{j}+m_{k}},\\
&\boldsymbol R_{i}=\frac{m_{i}\textbf{r}_{i}+m_{j}\textbf{r}_{j}+m_{k}\textbf{r}_{k}}{m_{i}+m_{j}+m_{k}}\equiv \mathbf{0},
\label{e13}
\end{aligned}
\end{eqnarray}
where $\{i$, $j$, $k\}$ = $\{$1, 2, 3$\}$, $\{$2, 3, 1$\}$ or $\{$3, 1, 2$\}$. $\textbf{r}_{i}$ and $m_{i}$ denote the position vector and the mass of the $i$th quark, respectively. $\textbf{\emph{R}}_{i}\equiv \textbf{0}$ means that the kinetic energy of the center of mass is not considered. Specially, for the JC-3 in Fig.~\ref{f1}, the following definitions are used in this work, $\boldsymbol\rho_{3}\equiv \boldsymbol\rho$ and $\boldsymbol\lambda_{3}\equiv \boldsymbol\lambda$.

Based on the above discussion and the heavy quark effective theory (HQET)~\cite{F303,F304,F403}, the spin and orbital wave function of a baryon state is assumed to have the coupling scheme
\begin{eqnarray}
|(J^{P})_{j},L\rangle = |\{[(l_{\rho} l_{\lambda} )_{L}(s_{1}s_{2})_{s_{12}}]_{j} s_{3}\}_{J }\rangle,
\end{eqnarray}
with $P=(-1)^{l_{\rho}+l_{\lambda}}$.
$l_{\rho}$($l_{\lambda}$), $L$ and $s_{12}$ are the quantum numbers of the relative orbital angular momentum $\textbf{\emph{l}}_{\rho}$ ($\textbf{\emph{l}}_{\lambda}$), total orbital angular momentum $\textbf{\emph{L}}$, and total spin of the light-quark pair $\mathbf{s}_{12}$, respectively. $j$ denotes the quantum number of the coupled angular momentum of $\textbf{\emph{L}}$ and $\textbf{s}_{12}$, so that the total angular momentum $J=j\pm\frac{1}{2}$.
More precisely, the baryon state is labeled with $(l_{\rho},l_{\lambda})nL(J^{P})_{j}$, in which $n$ is the quantum number of the radial excitation. It shows that such labeling of quantum states is acceptable, especially, $L$ being approximated as a good quantum number~\cite{F501}.
For the $\Sigma_{Q}$, $\Xi_{Q}^{'}$ and $\Omega_{Q}$ baryon families, $(-1)^{l_{\rho}+s_{12}}=-1$ should be also guaranteed due to the total antisymmetry of the wave function of the two light quarks, but  $(-1)^{l_{\rho}+s_{12}}=1$ for the $\Lambda_{Q}$ and $\Xi_{Q}$ families. All the conventions are based on the JC-3 in Fig.~\ref{f1}.

\subsection{Evaluations of the matrix elements}\label{sec2.3}

Since the orbital excited state $|\{[(l_{\rho} l_{\lambda} )_{L}(s_{1}s_{2})_{s_{12}}]_{j} s_{3}\}_{J }\rangle\equiv|\alpha\rangle_{3}$ is defined in the JC-3 as discussed above, the matrix elements of the Hamiltonian should be evaluated with the wave function $|\alpha\rangle_{3}$ of the Jacobi coordinates ($\boldsymbol\rho_{3}$, $\boldsymbol\lambda_{3}$). Here, the subscript 3 stands for JC-3. For a given orbital excited state $|\alpha\rangle_{3}$, the set of Gaussian basis functions {$|(\tilde{n},\alpha)^{G}_{3}\rangle$} form a set of finite-dimensional, non-orthogonal, and complete bases in a finite coordinate (radial) space, which are used in this work to achieve the high precision calculations of the matrix elements. This is the so-call Gaussian expansion method (GEM)~\cite{F602}. For the evaluation of the matrix element $\langle(\tilde{n},\alpha)^{G}_{3}|\hat{H}_{ij}|(\tilde{n}',\alpha)^{G}_{3}\rangle$ with $\hat{H}_{ij}(r_{ij})=\hat{H}(\rho_{k})$ ($k$=1, 2, 3 corresponds to JC-1, -2, -3, respectively), the Jacobi coordinates transformation needs to be performed as $\{\boldsymbol\rho_{3}$, $\boldsymbol\lambda_{3}\}$ $\rightarrow$ $\{\boldsymbol\rho_{k}$, $\boldsymbol\lambda_{k}\}$. However, it will be very tedious in the framework of the GEM.

This laborious process can be simplified by introducing the infinitesimally-shifted Gaussian (ISG) basis functions~\cite{F602}. With the help of the ISG basis functions, the matrix elements of the Hamiltonian terms $H_{0}$, $G_{ij}'$, $\tilde{S}_{ij}$, $\tilde{H}^{tensor}_{ij}$, and $\tilde{H}^{c}_{ij}$ can be evaluated rigorously in our previous works.  The GEM and ISG basis functions are briefly introduced in Appendix A and Appendix B, respectively. The detailed results can be found in Ref.~\cite{F502}.

In this work, the rigorous calculation of the spin-orbit terms $\langle(\tilde{n},\alpha)^{G}_{3}|\tilde{H}^{SO}_{ij}|(\tilde{n}',\alpha)^{G}_{3}\rangle$ is realized in the framework of the GEM and the ISG basis functions, by ignoring the mixing between different excited states. The detailed analysis is presented in Appendix C.

Now, all of the Hamiltonian matrix elements are evaluated. The eigenvalues of the Hamiltonian can be obtained rigorously, for the orbital excited states and their radial excited states.

\section{Results and discussions}\label{sec3}
For the $L$-wave excitation with $\textbf{\emph{L}}$=$\textbf{\emph{l}}_{\rho}$+$\textbf{\emph{l}}_{\lambda}$, there are an infinite number of orbital excitation modes.
Taking $L=1$ as an example, the excitation modes $(l_{\rho},l_{\lambda})_{L}$ are $(1,0)_{1}$, $(0,1)_{1}$, $(1,1)_{1}$, $(2,1)_{1}$, $(1,2)_{1}$, $(2,2)_{1}$, and so on. We assume that the excitation mode with the lowest energy level is the most stable and has the greatest probability of being observed experimentally, which dominates the structure of the excitation spectrum. This assumption is summarized as the HQD approximation (or the HQD mechanism)~\cite{F501}.

In the HQD mechanism, the orbital excited states of the singly heavy baryons mainly come from the $\lambda$-modes $(l_{\rho}=0,l_{\lambda})_{L=l_{\lambda}}$. But for the $P$-wave orbital excitations of the charm baryons with the $\mathbf{6}_{F}$ sector, i.e., the $\Sigma_{c}$, $\Xi'_{c}$ and $\Omega_{c}$ families, the HQD mechanism is broken because the mass of $c$ quark is not heavy enough, where both the $\lambda$-mode $(0,1)_{1}$ and the $\rho$-mode $(1,0)_{1}$ appear in their $P$-wave states.

Based on the above analyses, the $S$-, $P$- and $D$-wave states together with their radial excitations of the singly heavy baryons are investigated systematically, and the complete mass spectra are obtained. Taking the $\Lambda_{c}$ and the $\Sigma_{c}$ as examples, the contribution of each Hamiltonian term to the energy levels is given in Table~\ref{tta01} of Appendix D, so as to figure out the energy level splitting, the energy level evolution with each Hamiltonian term, and the formation of the fine structures. For the low-lying states, i.e., the $1S$-, $2S$-, $3S$-, $1P$-, $2P$- (only for the $\mathbf{\bar{3}}_{F}$ sector) and $1D$-wave states in this paper, their mass values and the root-mean-square radii are listed in Tables~\ref{tta1}-\ref{tta4} of Appendix D, and the corresponding mass spectra are presented in Fig.~\ref{f2}.

\subsection{Structure properties of singly heavy baryon spectra }\label{sec3.1}
(1) Contribution of each Hamiltonian term.

In these Hamiltonian terms, $\langle H_{mode}\rangle\equiv\langle H_{0}+H^{conf}\rangle$ depends on the excitation modes ($l_{\rho}, l_{\lambda}$) and dominates the main part of the energy levels. The other terms affect the shift and splitting of the energy levels. It is clearly displayed in Table~\ref{tta01}. As is shown in Table~\ref{tta01}, the tensor terms have little influence on the energy levels. The contact term $\langle H^{c}_{12}\rangle$ causes a big shift of the energy levels, nevertheless, has little effect on the energy level splitting. For the $\Sigma_{c}$ baryons, the contribution of the contact term $\langle H^{c}_{23(31)}\rangle$ to the energy level splitting decreases by orders of magnitude with the increase of $L$.

For the spin-orbit terms, $\langle H^{SO(v)}_{12}\rangle$ and $\langle H^{SO(s)}_{12}\rangle$ are equal to 0. The reason lies in that they are only related with $l_{\rho}$. In the (0,1) and (0,2) excitation modes ($l_{\rho}=0$), $\langle H^{SO(v)}_{12}\rangle$ and $\langle H^{SO(s)}_{12}\rangle$ vanish. While in the (1,0) mode ($l_{\rho}=1$) of the $\Sigma_{c}$ baryons, they are still equal to zero due to $s_{12}=0$ here, which is constrained by the condition $(-1)^{l_{\rho}+s_{12}}=-1$. So, the contribution of the spin-orbit terms comes only from the $\langle H^{SO}_{23}\rangle$ and the $\langle H^{SO}_{31}\rangle$. From Table~\ref{tta01}, one can see the $\langle H^{SO(v)}_{23(31)}\rangle$ and the $\langle H^{SO(s)}_{23(31)}\rangle$ always partially cancel each other out. But, they jointly lead to the shift and splitting of the energy levels. Especially, in the (1,0) mode, they cause a big splitting of the energy levels, which makes the $\mathbf{(1,0)}1P(\frac{1}{2}^{-})_{1}$ state intrude into the region of the $(0,1)1P$ states.

For the energy level splitting, the contribution of the spin-orbit terms is bigger than that of the contact terms. So, the spin-orbit interaction is very important for the excitation spectra structure of the singly heavy baryons.

(2) Heavy-quark dominance.

The HQD mechanism and its breaking in the orbital excitation of the heavy baryons were proposed and investigated in Refs.~\cite{F501,F507}, and the HQD mechanism dominates the structure of the excitation spectra. This mechanism means that the excitation mode with lower energy levels is always associated with the heavy quark(s), and the splitting of the energy levels is suppressed by the heavy quark(s) as well. In other words, the heavy quarks dominate the orbital excitation of singly and doubly heavy baryons, and determine the structures of their excitation spectra. The HQD mechanism is generally effective. But for the $1P$-wave orbital excitation of the singly charm baryons, it is slightly broken, since $c$ quark is not heavy
enough. From Tables~\ref{tta1}-\ref{tta4}, the results show that the mechanism holds up well under the rigorous calculation.

(3) Fine structures.

\begin{figure*}[p]
\centering
\includegraphics[trim=1cm 7cm 6cm 3cm, angle=270]{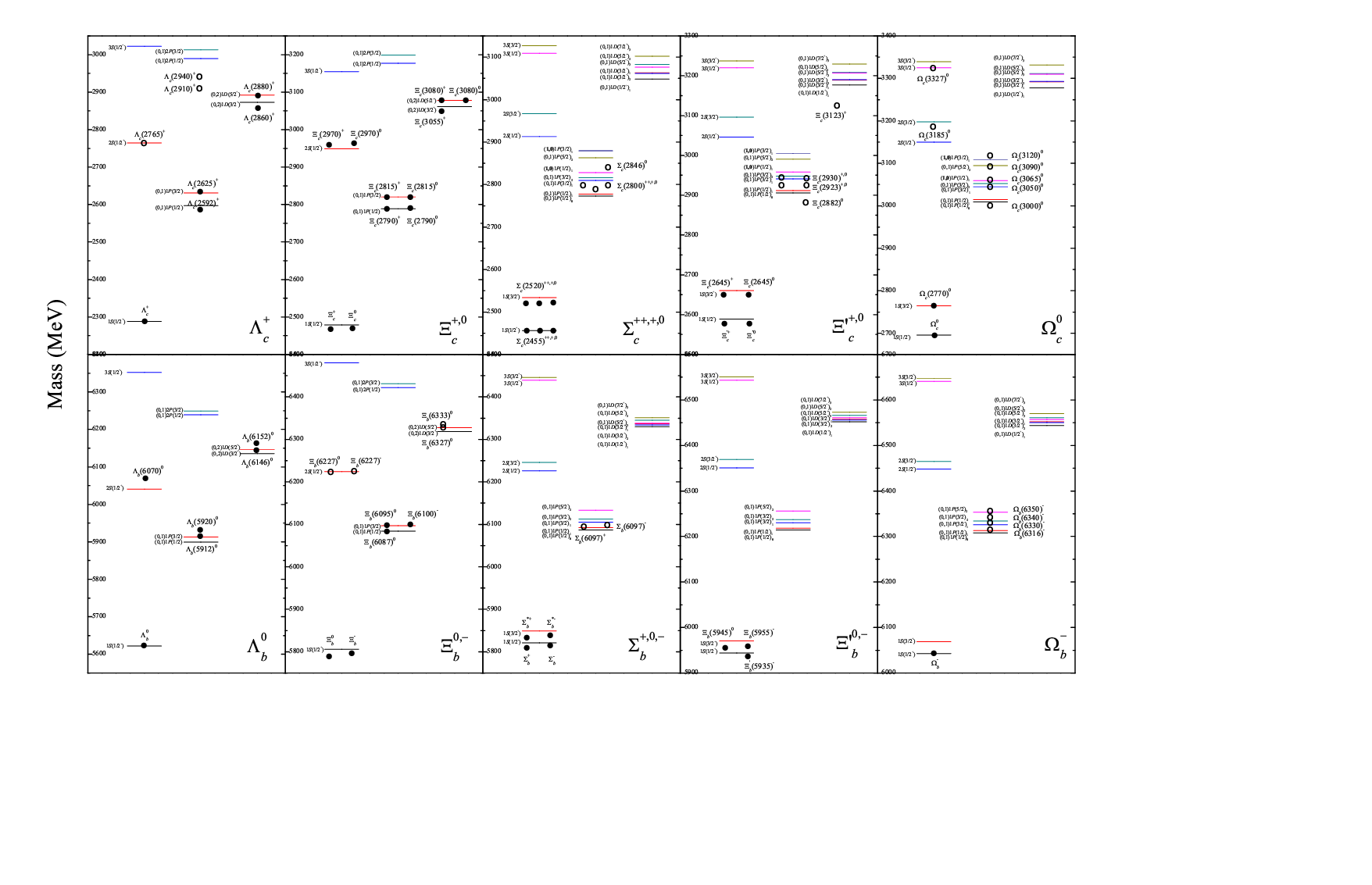}
\caption{Calculated spectra of the singly heavy baryons and the relevant experimental data~\cite{F201,F209,LHCb25}. `++',`+', `0' and `-' in the brackets indicate the charged states of baryons. The solid black circles denote the baryons with confirmed spin-parity values, and the open circles are the ones whose spin-parities have not been identified. }
\label{f2}
\end{figure*}

As is shown in Tables~\ref{tta2}-\ref{tta4} and Fig.~\ref{f2}, the rigorous calculation reveals the perfect fine structures of the excitation spectra, not only for all the $1P$-wave states, but also for the $1D$-wave states of the charm baryons $\Sigma_{c}$, $\Xi_{c}'$ and $\Omega_{c}$. According to the data of the $\Omega_{c}$ baryons, the fine structure of the $1P$-excited charm baryons ($\Sigma_{c}$, $\Xi'_{c}$ and $\Omega_{c}$) should be composed of the 5 energy levels which are the $(0,1)1P(\frac{1}{2}^{-})_{0,1}$, $(0,1)1P(\frac{3}{2}^{-})_{1,2}$, $(1,0)1P(\frac{1}{2}^{-})_{1}$ (as an intrude state), $(0,1)1P(\frac{5}{2}^{-})_{2}$ and $(1,0)1P(\frac{3}{2}^{-})_{1}$, respectively. Based on the data of the $\Omega_{b}$ baryons, however, the fine structure of the $1P$-wave states of the bottom baryons ($\Sigma_{b}$, $\Xi'_{b}$ and $\Omega_{b}$) may contain the 4 energy levels, they are the $(0,1)1P(\frac{1}{2}^{-})_{0,1}$, $(0,1)1P(\frac{3}{2}^{-})_{1}$, $(0,1)1P(\frac{3}{2}^{-})_{2}$ and $(0,1)1P(\frac{5}{2}^{-})_{2}$, respectively. For the $1D$-wave states of the $\Sigma_{c}$, $\Xi'_{c}$ and $\Omega_{c}$ baryons, there are clear and distinct 4 energy levels as shown in Fig.~\ref{f2}. The predicted fine structure of the $1D$-wave states has yet to be confirmed by the future experiments.

(4) Missing states.

In the relativized quark model, the calculations in Refs.~\cite{F402,Fp003} predicted a substantial number of `missing' states, compared to the experimental observations of the singly heavy baryons. The practice of reducing the internal degrees of freedom, such as the heavy quark-light diquark picture~\cite{F405}, predicted significantly fewer states than the former, however, lacks a reasonable physical explanation~\cite{F101,zhao24}. Now, under the HQD mechanism, the rigorous calculation can reproduce well the data, and the problem of the missing states disappears thereof. So, the HQD mechanism in the genuine three-body picture might be a natural solution to the missing states.

(5) Clustering effect.

The heavy quark-light diquark picture achieved great successes in describing the spectra of the singly heavy baryons, based on an important concept of the `diquark' or the quark cluster~\cite{F405}. By taking account of the contribution of the quark cluster, the fine structure was preliminarily explained in our previous work~\cite{F507}, which hints that there might be the clustering effect inside a singly heavy baryon. Now, the rigorous calculation shows that, without introducing the concept of the `diquark' or the quark cluster, the excitation spectra and their fine structures can also be reproduced very well. So, there is no indication that the clustering effect is indispensable inside a singly heavy baryon.

(6) Spin-orbit terms.

In both the light-quark baryons and the heavy-quark baryons, the treatment of the spin-orbit terms used to be a difficult problem~\cite{F306,Fp001,Fp002}. This is mainly due to the following two reasons. One is that the experimental data were not sufficient, and the other is that the rigorous model calculation was difficult. Both difficulties have now been overcome in the research of the singly heavy baryons, i.e., there are enough experimental data currently and the rigorous calculations has been implemented. Table~\ref{tta01} shows clearly the contribution of each spin-orbit term, which demonstrates its irreplaceable role in accurately reproducing the fine structures. And an earlier assertion is confirmed here, namely, the contribution of the spin-orbit terms must indeed be fully considered before the fine structures can be well explained in the singly heavy baryon spectra~\cite{F306}. Therefore, based on this study, it is concluded that the spin-orbit terms of the relativized quark model are reasonable for describing the singly heavy baryon spectra, and the `spin-orbit puzzle'~\cite{F306,Fp001,Fp002} does not exist anymore here. Note that this work ignores the mixing between different excited states, whose effect on the energy levels still needs to be further studied.

\subsection{Excitation spectra and experimental data}\label{sec3.2}

In our previous works, the assignments of the observed baryons have been discussed, and a detailed comparison of our results with other theoretical estimations has been presented as well~\cite{F501,F502,F503,F507}. In this work, the rigorous calculation mainly improves the results of the fine structure. So, the following discussion will focus on the systematic analysis of the model calculations, by comparing the predicted excitation spectra with the experimental data.

All of the observed masses of the singly heavy baryons and the predicted spectra are plotted together in Fig.~\ref{f2}. The detailed experimental data and calculated results are listed in Tables~\ref{tta1}-\ref{tta4}, for the $\Lambda_{c(b))}$, $\Xi_{c(b)}$, $\Sigma_{c(b)}$, $\Xi'_{c(b)}$ and $\Omega_{c(b)}$ baryons, respectively.
As is shown in Fig.~\ref{f2} and Tables~\ref{tta1}-\ref{tta4}, most of the observed masses match well with the predicted spectra, and the maximum deviation between the calculated masses and the data is generally not more than 20 MeV.

(1) $\Lambda_{c(b)}$ and $\Xi_{c(b)}$ baryons.

 The $\Lambda_{c(b)}$ and $\Xi_{c(b)}$ baryons belong to the $\mathbf{\bar{3}}_{F}$ sector.  They have the same spectral structure. Fig.~\ref{f2} shows that the match between the calculation results and the data is good on the whole, except for the $\Lambda_{c}(2910)^{+}$ and the $\Lambda_{c}(2940)^{+}$. The $\Lambda_{c}(2940)^{+}$ was measured by the LHCb collaboration in 2017~\cite{P2940}, and a narrow peak was seen in $pD^{0}$ and in $\Lambda_{c}^{+}\pi^{+}\pi^{-}$. It was not seen in $pD^{+}$, and therefore it might be a $\Lambda_{c}^{+}$ baryon. Its $J^{P}=3/2^{-}$ is favored, but not certain~\cite{F201}. The $\Lambda_{c}(2910)^{+}$ was reported by the Belle collaboration in 2022~\cite{P2910}. It was considered as the candidate of the heavy quark symmetry doublet partner to the $\Lambda_{c}(2940)^{+}$~\cite{F201}. In Fig.~\ref{f2}, one can see these two baryons have to be assigned as the $2P$-doublet states, if they belong to the $\Lambda_{c}$ family. However, the difference between their measured masses and predicted ones is so big that it is far beyond the allowable error range of the theoretical calculation. So, the $\Lambda_{c}(2910)^{+}$ and $\Lambda_{c}(2940)^{+}$  are probably not the members of the $\Lambda_{c}$ family. In some theoretical studies, they were considered as the molecular states~\cite{F315,F315p}. If only their mass values are considered, however, they are more like the candidates of the $2S$-doublet states in the $\Sigma_{c}$ family as shown in Fig.~\ref{f2} and Table~\ref{tta2}. It needs to be further confirmed by experiments.

The $\Xi_{b}(6227)^{0,-}$ baryons were measured precisely by the LHCb collaboration in 2021~\cite{F213}, but their $J^{P}$ values remain unconfirmed. According to their mass values, the $\Xi_{b}(6227)^{0,-}$ baryons could be assigned as the $2S(\frac{1}{2}^{+})$ state of the $\Xi_{b}$ family as shown in Fig.~\ref{f2}. Alternatively, they might be the candidates of the $1P(\frac{1}{2}^{-})_{0,1}$ state or the $1P(\frac{3}{2}^{-})_{1}$ state of the $\Xi'_{b}$ family.

(2) $\Sigma_{c}$ and $\Sigma_{b}$ baryons.

 The $\Sigma_{c}(2800)^{++,+,0}$ baryons were reported by the Belle Collaboration in 2005~\cite{F210}. The $\Sigma_{c}(2846)^{0}$ was observed by the BaBar collaboration, with $m=2846\pm8\pm10$ MeV~\cite{F209}, which has not been collected by the PDG so far. In this work, it is assumed to be a real baryon. Based on the calculation, the $\Sigma_{c}(2846)^{0}$ and the $\Sigma_{c}(2800)^{++,+,0}$ are in the region of the $1P$-wave states. By examining their mass values and the fine structure of the $1P$-wave states shown in Fig.~\ref{f2}, the $\Sigma_{c}(2800)^{++,+,0}$ could be assigned as the $(0,1)1P(\frac{1}{2}^{-})_{0,1}$ states, and the $\Sigma_{c}(2846)^{0}$ could be considered as the intrude state $(1,0)1P(\frac{1}{2}^{-})_{1}$.

 The case of the $\Sigma_{b}(6097)^{+,-}$ is similar to that of the $\Sigma_{c}(2800)^{++,+,0}$. So, we can safely conclude that the $J^{P}$ of the $\Sigma_{b}(6097)^{+,-}$ is likely to be $\frac{1}{2}^{-}$. And they should be the $(0,1)1P(\frac{1}{2}^{-})_{0,1}$ states.

(3) $\Xi_{c}^{'}$ and $\Xi_{b}^{'}$ baryons.

 A charged $\Xi_{c}(2930)^{+}$ baryon was observed by the Belle collaboration in 2018~\cite{F204}. Later, the $\Xi_{c}(2923)^{0}$, $\Xi_{c}(2939)^{0}$ and $\Xi_{c}(2964)^{0}$ states were observed with a large significance by the LHCb collaboration~\cite{F203}. Very recently, a new charmed baryon $\Xi_{c}(2923)^{+}$ was firstly observed by the LHCb collaboration~\cite{LHCb25}. In the new PDG data, these baryons were relabeled as the $\Xi_{c}(2923)^{0}$, $\Xi_{c}(2930)^{+,0}$ and $\Xi_{c}(2970)^{0}$.  The $\Xi_{c}(2970)^{0}$ and its isospin partner $\Xi_{c}(2970)^{+}$ are assigned as the $2S(\frac{1}{2}^{+})$ state of the $\Xi_{c}$ family~\cite{F201}. While the $\Xi_{c}(2882)^{0}$~\cite{P2882}, $\Xi_{c}(2923)^{+,0}$ and $\Xi_{c}(2930)^{+,0}$ exhibit the fine structure of the $1P$-wave states in the $\Xi'_{c}$ family. As is shown in Fig.~\ref{f2}, their assignments could be the $(0,1)1P(\frac{1}{2}^{-})_{0,1}$, $(0,1)1P(\frac{3}{2}^{-})_{1,2}$ and  $(1,0)1P(\frac{1}{2}^{-})_{1}$ states, respectively.

 The $\Xi_{c}(3123)^{+}$ was observed by the BaBar Collaboration in 2007~\cite{F205}. It is difficult to make a good assignment for the $\Xi_{c}(3123)^{+}$. As is shown in Fig.~\ref{f2}, we consider it as a candidate of the $1D$-wave state, even though its mass value is too small. Alternatively, it could be the $2S(\frac{3}{2}^{+})$ state.

 If we assume that the $\Xi_{b}(6227)^{0,-}$ baryons are the strange partner of the $\Sigma_{b}(6097)^{+,-}$, we find there are great similarities between them. So, the $\Xi_{b}(6227)^{0,-}$ baryons could also be assigned as the same states as the $\Sigma_{b}(6097)^{+,-}$, instead of the $2S(\frac{1}{2}^{+})$ state of the $\Xi_{b}$ family as mentioned above.

(4) $\Omega_{c}$ and $\Omega_{b}$ baryons.

For these two families, the predicted fine structures of the $1P$-wave states reproduce the data perfectly, as shown in Fig.~\ref{f2}. Their assignments are listed in Table~\ref{tta4}. The $\Omega_{c}(3185)^{0}$ is likely to be the $2S(\frac{3}{2}^{+})$ state. The $\Omega_{c}(3327)^{0}$ is assigned as the $3S(\frac{1}{2}^{+})$ state, but its mass value overlaps with those of the $1D$-wave states.

(5) Baryons in the fine structures.

The $\Sigma_{c}(2800)^{++,+,0}$, $\Sigma_{c}(2846)^{0}$ and $\Sigma_{b}(6097)^{+,-}$ have a common feature, i.e., their decay widths are much more than 15 MeV. For the $\Xi'_{c}$, $\Omega_{c}$ and $\Omega_{b}$ baryons in the fine structures, however, their decay widths are overall smaller than 15 MeV. Given the similarity in the spectral structure of these $\Sigma_{c(b)}$, $\Xi'_{c(b)}$ and $\Omega_{c(b)}$ families, it may be true that the decay widths of the baryons in the fine structures could all be small.
From this point of view, the $\Sigma_{c}(2800)^{++,+,0}$, $\Sigma_{c}(2846)^{0}$ and $\Sigma_{b}(6097)^{+,-}$ might be the superpositions of several quantum states, and that more precise measurements may reveal their fine structures further. The $\Xi_{b}(6227)^{0,-}$ would have the same problem if they belong to the $\Xi'_{b}$ family, as well as the assignment of the $\Omega_{c}(3327)^{0}$ as mentioned above.

In Ref.~\cite{F4071}, the following chain was found by analyzing the universal behavior of the mass gaps of the baryons,
 \begin{eqnarray}
\begin{aligned}
\Sigma_{c}(2846)^{0}\leftrightarrow\Xi_{c}^{'}(2964)^{0}\leftrightarrow\Omega_{c}(3090)^{0} ,
\end{aligned}
\end{eqnarray}
which implies that these baryons are in the same quantum state. Now, the $\Xi_{c}(2964)^{0}$ (relabeled as $\Xi_{c}(2970)^{0}$) has been considered as the member of the $\Xi_{c}$ family. As is shown in Fig.~\ref{f2}, the updated chain should be as follow,
\begin{eqnarray}
\begin{aligned}
\Sigma_{c}(2846)^{0}\leftrightarrow\Xi'_{c}(2930)^{0}\leftrightarrow\Omega_{c}(3065)^{0},
\end{aligned}
\end{eqnarray}
if the $\Sigma_{c}(2846)^{0}$ is a single state.

\subsection{Reliability of the model }\label{sec3.3}

Some approximate calculations were adopted in our previous works actually. In Refs.~\cite{F502,F503}, the $\tilde{H}^{hyp}_{13}$ and $\tilde{H}^{hyp}_{23}$ terms were ignored in the hyperfine interaction. The spin-orbit interaction only contained the $\tilde{H}^{SO}_{12}$ term coming from the light quark pair and a part of the $\tilde{H}^{SO}_{d-Q}$ term contributed jointly by the heavy quark ($Q$) and the light-diquark ($d$) (only including the leading order contribution as the Eq.(33) in Ref.~\cite{Fp002}). In Ref.~\cite{F507}, the light-diquark approximation was considered completely, where the hyperfine interaction was represented by the $\tilde{H}^{hyp}_{12}$ and $\tilde{H}^{hyp}_{d-Q}$ terms, and the spin-orbit interaction contained the $\tilde{H}^{SO}_{12}$ and $\tilde{H}^{SO}_{d-Q}$ terms. In this work, all of the Hamiltonian terms are obtained without approximation. As a result, most of the energy levels of the excited states in this work are shifted, even for some of the $S$-wave radial excited states, compared to those in our previous works.

Since the parameters used in the present work are given without any uncertainty, they certainly do not result in any uncertainty in the calculated results. We here evaluate the deviations of the calculated masses of the 74 baryons from the measured ones as shown in Table~\ref{a8}. Most of the deviations are less than 20 MeV. And the arithmetic average deviation is less than 10 MeV, which is consistent with the estimation result in Ref.~\cite{F401}.

As is shown in Fig.~\ref{f2} and Table~\ref{a8}, the predicted mass spectra in this work can reproduce the data nicely on the whole, for all the singly heavy baryon families. The shell structure of the spectra is clearly shown. It implies that this model can successfully describe the singly heavy baryon spectra without approximation.

The fine structures can be reproduced well, especially for the $\Omega_{c}$ and $\Omega_{b}$ families. It shows the rationality of the Hamiltonian based on the two-body interactions of the relativized quark model.

While, for the excitation spectrum of each family, there is a little systematic deviation between the predicted mass values and the data. For a few baryons, such as the $\Xi_{c}(3123)^{+}$, the theoretical results cannot explain the data reasonably. So, some improvements of this model should be tried, such as a parameter optimization.

In summary, under the HQD mechanism, the relativized quark model can describe the excitation spectra and the fine structures correctly. Based on the relativized quark model, the method used in this work should be reliable in the research of the singly heavy baryons spectroscopy.

\section{Conclusions}\label{sec4}

In this work, the rigorous calculation of the spin-orbit terms of the relativized quark model is realized based on the GEM and the ISG basis functions, by ignoring the mixing between different excited states. Then, the complete mass spectra of the singly heavy baryons are obtained rigorously in the framework of the relativized quark model and under the HQD mechanism. On these bases, the systematical analyses are carried out for the reliability and predictive power of the model, the fine structure of the singly heavy baryon spectra, the assignments of the excited baryons, and some important topics about the heavy baryon spectroscopy, such as the missing states, the clustering effect, the `spin-orbit puzzle', etc.

The main work done and main results obtained in the present paper are as follows:

(1) The contribution of each Hamiltonian term to the energy levels is figured out.

(2) The HQD mechanism is further confirmed.

(3) The fine structures of the singly heavy baryons are presented.

(4) The missing states in the singly heavy baryon spectra disappear naturally under the HQD mechanism.

(5) There is no indication that the clustering effect is indispensable in a singly heavy baryon.

(6) The spin-orbit terms of the relativized quark model are reasonable for describing the singly
heavy baryon spectra, and the `spin-orbit puzzle' does not exist here.

(7) The $\Lambda_{c}(2910)^{+}$ and $\Lambda_{c}(2940)^{+}$  are probably not the members of the $\Lambda_{c}$ family. While, they are more like the candidates of the $2S$-doublet states in the $\Sigma_{c}$ family, if only their mass values are considered.

(8) It is difficult to make a good assignment for the $\Xi_{c}(3123)^{+}$ in this work.

(9) The $\Sigma_{c}(2800)^{++,+,0}$, $\Sigma_{c}(2846)^{0}$ and $\Sigma_{b}(6097)^{+,-}$ may not be single states, and more precise measurements are advised for uncovering their fine structures further.

In summary, the rigorous calculation shows that under the HQD mechanism, the relativized quark model can describe the excitation spectra and the fine structures of the singly heavy baryons correctly and precisely. Based on the relativized quark model, the method used in this work should be reliable in the research of the singly heavy baryons spectroscopy. And some improvements of this method should be tried later, for a deep understanding of the properties of the singly heavy baryon spectroscopy and the strong interaction in the non-perturbative regime of QCD.

\begin{large}
\section*{Acknowledgements}
\end{large}

We thank the reviewers for their valuable comments and suggestions. JG would like to thank Professor Yinsheng Ling from Soochow University for his helpful discussion and great encouragement.
This research was supported by the Open Project of Guangxi Key Lab of Nuclear Physics and Technology (No. NLK2023-04), the Central Government Guidance Funds for Local Scientific and Technological Development in China (No. Guike ZY22096024), the Natural Science Foundation of Guizhou Province-ZK[2024](General Project)650, the National Natural Science Foundation of China (Grant Nos. 11675265, 12175068), the Continuous Basic Scientific Research Project (Grant No. WDJC-2019-13) and the Leading Innovation Project (Grant No. LC 192209000701).

\begin{large}
\section*{Appendices }
\end{large}
\subsection{Gaussian expansion method (GEM)}\label{A}
Given a set of the orbital quantum numbers $\{l$, $m\}$, the Gaussian basis function $|(nlm)^{G} \rangle$ is commonly written in  position space as
 \begin{eqnarray}
\begin{aligned}
 &\phi^{G}_{nlm}(\textbf{r})=\phi^{G}_{nl}(r)Y_{lm}(\hat{\textbf{r}}),\\
 &\phi^{G}_{nl}(r)=N_{nl}r^{l}e^{-\nu_{n}r^{2}},\\
 &N_{nl}=\sqrt{\frac{2^{l+2}(2\nu_{n})^{l+3/2}}{\sqrt{\pi}(2l+1)!!}},
\end{aligned}
\end{eqnarray}
with
\begin{eqnarray}
\begin{aligned}
& \nu_{n}=\frac{1}{r^{2}_{n}},\\
& r_{n}=r_{1}a^{n-1}\ \ \ (n=1,\ 2,\ ...,\ n_{max}).
\end{aligned}
\end{eqnarray}
$\{r_{1}, a, n_{max}\}$ (or equivalently $\{n_{max},r_{1},r_{n_{max}}\}$) are the Gaussian size
parameters and commonly related to the scale in question~\cite{F602}.  The optimized values of $\{n_{max}=10$, $r_{1}=0.18$ GeV$^{-1}$, $r_{n_{max}}=15$ GeV$^{-1}\}$ are finally selected for the heavy baryons in this work. Details can be found in Refs.~\cite{F502,F503}.

The set $\{\phi^{G}_{nlm}\}$ forms a set of finite-dimensional, non-orthogonal, and complete bases,
 \begin{eqnarray}
\begin{aligned}
 &N_{n,n'}=\langle\phi^{G}_{nlm}|\phi^{G}_{n'lm}\rangle=(\frac{2\sqrt{\nu_{n}\nu_{n'}}}{\nu_{n}+\nu_{n'}})^{l+\frac{3}{2}},\\
 &1=\sum^{n_{max}}_{n=1}\sum^{n_{max}}_{n'=1}|\phi^{G}_{nlm}\rangle(N^{-1})_{nn'}\langle\phi^{G}_{n'lm}|.
 \end{aligned}
\end{eqnarray}
An arbitrary wave function $\psi_{lm}(\mathbf{r})$ can be expended in a set of definite orbital quantum states,
 \begin{eqnarray}
\begin{aligned}
 &|\psi_{lm}\rangle=\sum^{n_{max}}_{n,n'=1}|\phi^{G}_{nlm}\rangle(N^{-1})_{nn'}\langle\phi^{G}_{n'lm}|\psi_{lm}\rangle
 \equiv\sum^{n_{max}}_{n=1}C_{n}|\phi^{G}_{nlm}\rangle.
 \end{aligned}
\end{eqnarray}

In the definite orbital quantum state, the matrix element of an operator $\hat{O}$ reads,
\begin{eqnarray}
\begin{aligned}
O_{nn'}=\langle\phi^{G}_{nlm}|\hat{O}|\phi^{G}_{n'lm}\rangle.
\end{aligned}
\end{eqnarray}

Given $|(nlm)^{G}\rangle\equiv|n\rangle$ and $|(n'lm)^{G}\rangle\equiv|n'\rangle$, and operators $\hat{O}_{1}$, $\hat{O}_{2}$ and $\hat{O}_{3}$, the matrix element of their inner product in the set of bases is expressed as,
\begin{eqnarray}
\notag
&&\langle n|\hat{O}_{1}\hat{O}_{2}\hat{O}_{3}|n'\rangle\\
\notag
&&=\sum_{\{n_{i},n'_{i}\}}\langle n|\hat{O}_{1}|n_{1}\rangle(N^{-1})_{n_{1}n_{1}'}\langle n_{1}'|\hat{O}_{2}|n_{2}\rangle
(N^{-1})_{n_{2}n_{2}'}\langle n_{2}'|\hat{O}_{3}|n'\rangle\\
&&=\sum_{\{n_{i},n'_{i}\}}(O_{1})_{nn_{1}}(N^{-1})_{n_{1}n_{1}'}(O_{2})_{n_{1}'n_{2}}(N^{-1})_{n_{2}n_{2}'}(O_{3})_{n_{2}'n'}.
\end{eqnarray}
Here, $\sum_{\{n_{i},n'_{i}\}}$ means sum over all the intermediate indices. The expectation value of an operator $\hat{O}$ in a state $|\alpha\rangle$ is written as,
\begin{eqnarray}
\notag
\frac{\langle\alpha|\hat{O}|\alpha\rangle}{\langle\alpha|\alpha\rangle}
&&=\frac{\sum_{\{n\}}\langle\alpha|n^{G}_{1}\rangle(N^{-1})_{n_{1}n_{1}'}\langle n_{1}'|\hat{O}|n^{G}_{2}\rangle(N^{-1})_{n_{2}n_{2}'}\langle n'^{G}_{2}|\alpha\rangle}{\sum_{\{n\}}\langle\alpha|n^{G}_{3}\rangle(N^{-1})_{n_{3}n_{3}'}\langle n'^{G}_{3}|n^{G}_{4}\rangle(N^{-1})_{n_{4}n_{4}'}\langle n'^{G}_{4}|\alpha\rangle}\\
&&=\frac{\sum_{\{n\}}C^{*}_{n_{1}'}O_{n_{1}'n_{2}}C_{n_{2}}}{\sum_{\{n\}}C^{*}_{n_{3}'}N_{n'_{3}n_{4}}C_{n_{4}}},
\end{eqnarray}
in the set of the Gaussian bases.

Now, given a definite quantum state $|(ls)_{JM_{J}}\rangle$, the generalized Gaussian basis function ($|[n,(ls)_{JM_{J}}]^{G}\rangle$) is commonly written as
\begin{eqnarray}
\begin{aligned}
 &|[n,(ls)_{JM_{J}}]^{G}\rangle=\sum_{m_{l},m_{s}}(lm_{l}sm_{s}|JM_{J})\times|(nlm_{l})^{G}\rangle\otimes|sm_{s}\rangle.
 \end{aligned}
\end{eqnarray}
The set $\{|[n,(ls)_{JM_{J}}]^{G}\rangle\}$ also forms a set of finite-dimensional, non-orthogonal, and complete bases,
 \begin{eqnarray}
\begin{aligned}
 &N_{n,n'}=\langle [n,(ls)_{JM_{J}}]^{G}|[n',(ls)_{JM_{J}}]^{G}\rangle=(\frac{2\sqrt{\nu_{n}\nu_{n'}}}{\nu_{n}+\nu_{n'}})^{l+\frac{3}{2}},\\
 &1=\sum^{n_{max}}_{n=1}\sum^{n_{max}}_{n'=1}|[n,(ls)_{JM_{J}}]^{G}\rangle(N^{-1})_{nn'}\langle [n',(ls)_{JM_{J}}]^{G}|.
 \end{aligned}
\end{eqnarray}

For a singly heavy baryon, we introduce two independent sets of the Gaussian basis functions $|(n_{\rho}l_{\rho}m_{\rho})^{G} \rangle$ and $|(n_{\lambda}l_{\lambda}m_{\lambda})^{G} \rangle$ based on the JC-3 in Fig.~\ref{f1}. Given a definite quantum state $|\{[(l_{\rho} l_{\lambda} )_{L}(s_{1}s_{2})_{s_{12}}]_{j} s_{3}\}_{JM_{J} }\rangle\equiv|\alpha\rangle_{3}$ (corresponding to the JC-3), the generalized Gaussian basis function has the form below,
\begin{eqnarray}
\notag
|(\tilde{n},\alpha)^{G}_{3}\rangle &&=\sum_{\{m_{\xi}\}}\{CG_{\xi}\}\times|(n_{\rho}l_{\rho}m_{\rho})^{G} \rangle\otimes|(n_{\lambda}l_{\lambda}m_{\lambda})^{G} \rangle\\
&&\otimes|s_{1}m_{s_{1}}\rangle\otimes|s_{2}m_{s_{2}}\rangle\otimes|s_{3}m_{s_{3}}\rangle,
\end{eqnarray}
where $\{m_{\xi}\}$ denote all the 3rd components of the orbital angular momenta and spins, $\{CG_{\xi}\}$ are the products of all the C-G coefficients. $\tilde{n}$ is obtained by combining $n_{\rho}$ and $n_{\lambda}$, e.g., $\tilde{n}=(n_{\rho}-1)\times n_{max}+n_{\lambda}$ as $n_{\rho(\lambda)}=1,\cdot\cdot\cdot,n_{max}$.

The non-orthogonal and complete relations are as follows,
\begin{eqnarray}
\begin{aligned}
\notag
 &N_{\tilde{n},\tilde{n}'}=\langle (\tilde{n},\alpha)^{G}_{3}|(\tilde{n}',\alpha)^{G}_{3}\rangle=(\frac{2\sqrt{\nu_{n_{\rho}}\nu_{n_{\rho}'}}}{\nu_{n_{\rho}}+\nu_{n_{\rho}'}})^{l_{\rho}+\frac{3}{2}}
 \times (\frac{2\sqrt{\nu_{n_{\lambda}}\nu_{n_{\lambda}'}}}{\nu_{n_{\lambda}}+\nu_{n_{\lambda}'}})^{l_{\lambda}+\frac{3}{2}},\\
 \notag
 &1=\sum^{n^{2}_{max}}_{\tilde{n}=1}\sum^{n^{2}_{max}}_{\tilde{n}'=1}|(\tilde{n},\alpha)^{G}_{3}\rangle(N^{-1})_{\tilde{n}\tilde{n}'}\langle (\tilde{n}',\alpha)^{G}_{3}|.
 \end{aligned}
\end{eqnarray}

In the non-orthogonal representation of $|(\tilde{n}\alpha)^{G}_{3}\rangle$, the solution of the eigenenergy $E$ belongs to a generalized matrix eigenvalue problem
\begin{eqnarray}
\begin{aligned}
\sum^{n^{2}_{max}}_{\tilde{n}'=1}(H_{\tilde{n}\tilde{n}'}-EN_{\tilde{n}\tilde{n}'})C_{\tilde{n}'}=0.
\end{aligned}
\end{eqnarray}
The matrix element of an operator $\hat{H}$ reads,
\begin{eqnarray}
\notag
H_{\tilde{n}\tilde{n}'} &&=\langle(\tilde{n},\alpha)^{G}_{3}|\hat{H}|(\tilde{n}',\alpha)^{G}_{3}\rangle\\
\notag
&&=\sum_{\{m_{\xi}\},\{m_{\xi}'\}}\{CG_{\xi}\}\times\{CG_{\xi'}\}\times\langle(n_{\rho}l_{\rho}m_{\rho})^{G}|\langle(n_{\lambda}l_{\lambda}m_{\lambda})^{G}|\\
\notag
&& \times\langle s_{1}m_{s_{1}}|\langle s_{2}m_{s_{2}}|\langle s_{3}m_{s_{3}}|\hat{H}|s_{1}m_{s_{1}}'\rangle|s_{2}m_{s_{2}}'\rangle|s_{3}m_{s_{3}}'\rangle \\
\notag
&&\times|(n_{\rho}'l_{\rho}m_{\rho}')^{G} \rangle|(n_{\lambda}'l_{\lambda}m_{\lambda}')^{G} \rangle\\
&&\equiv\sum_{\{m_{\xi},m_{\xi}'\}}\{CG_{\xi}\times CG_{\xi'}\}\times H_{(n_{\rho}n_{\lambda},n_{\rho}'n_{\lambda}');(m_{s_{1,2,3}},m_{s_{1,2,3}}')}.
\end{eqnarray}
The matrix element evaluation of $H_{\tilde{n}\tilde{n}'}$ is finally implemented for $H_{(n_{\rho}n_{\lambda},n_{\rho}'n_{\lambda}');(m_{s_{1,2,3}},m_{s_{1,2,3}}')}$.
For the two-body interaction $\hat{V}_{ij}(r_{ij})$,
\begin{eqnarray}
\notag
&&[V_{ij}(r_{ij})]_{(n_{\rho}n_{\lambda},n_{\rho}'n_{\lambda}');(m_{s_{1,2,3}},m_{s_{1,2,3}}')}
 =[V(\rho_{k})]_{(n_{\rho}n_{\lambda},n_{\rho}'n_{\lambda}');(m_{s_{1,2,3}},m_{s_{1,2,3}}')}\\
\notag
&&\equiv\langle(n_{\rho_{3}}l_{\rho_{3}}m_{\rho_{3}})^{G}|\langle(n_{\lambda_{3}}l_{\lambda_{3}}m_{\lambda_{3}})^{G}|\\
\notag
&& \times\langle s_{1}m_{s_{1}}|\langle s_{2}m_{s_{2}}|\langle s_{3}m_{s_{3}}|\hat{V}(\rho_{k})|s_{1}m_{s_{1}}'\rangle|s_{2}m_{s_{2}}'\rangle|s_{3}m_{s_{3}}'\rangle \\
&&\times|(n_{\rho_{3}}'l_{\rho_{3}}m_{\rho_{3}}')^{G} \rangle|(n_{\lambda_{3}}'l_{\lambda_{3}}m_{\lambda_{3}}')^{G} \rangle.
\end{eqnarray}

If the matrix element $[V(\rho_{k})]_{(n_{\rho}n_{\lambda},n_{\rho}'n_{\lambda}');(m_{s_{1,2,3}},m_{s_{1,2,3}}')}$ is independent of the spin operator, it can be written further as
$[V(\rho_{k})]_{(n_{\rho}n_{\lambda},n_{\rho}'n_{\lambda}')}\delta_{m_{s1}m'_{s1}}\delta_{m_{s2}m'_{s2}}\delta_{m_{s3}m'_{s3}}$.
The matrix element $[V(\rho_{k})]_{(n_{\rho}n_{\lambda},n_{\rho}'n_{\lambda}')}$ can be calculated with the help of the Jacobi coordinates transformation $(\boldsymbol\rho_{3}$,$\boldsymbol\lambda_{3})$ $\rightarrow$ $(\boldsymbol\rho_{k}$,$\boldsymbol\lambda_{k})$ ($k$=1,~2,~3), but it will be very tedious in the framework of the GEM.

\subsection{Infinitesimally-shifted Gaussian (ISG) basis functions}\label{B}

In the calculation of Hamiltonian matrix elements of three-body systems, particularly, when the Jacobi coordinates transformations are employed, integrations over all of the radial and angular coordinates become laborious even with the Gaussian basis functions. This process can be simplified by introducing the infinitesimally-shifted Gaussian (ISG) basis functions by
\begin{eqnarray}
\notag
\phi_{nlm}^{G} &&=N_{nl}r^{l}e^{-\nu_{n}r^{2}}Y_{lm}(\mathbf{\hat{r}})\\
&&=N_{nl}\lim_{\varepsilon\rightarrow~0}\frac{1}{(\nu_{n}\varepsilon)^{l}}\sum _{\tilde{k}=1}^{\tilde{k}_{max}}C_{lm,\tilde{k}}e^{-\nu_{n}(\mathbf{r}-\varepsilon \mathbf{D}_{lm,\tilde{k}})^{2}},
\end{eqnarray}
where, $r^{l}Y_{lm}(\mathbf{\hat{r}})$ is replaced by a set of coefficients $C_{lm,\tilde{k}}$ and vectors $\mathbf{D}_{lm,\tilde{k}}$. In this way, the Jacobi coordinates transformation just needs to be completed in the exponent section.

Considering an arbitrary matrix element $[V(\rho_{k})]_{(n_{\rho}n_{\lambda},n_{\rho}'n_{\lambda}')}$, $V(\rho_{k})$ is a scalar function of the radii $\rho_{k}$ ($k=1,~2,~3$, corresponding to the JC-1, -2, -3, respectively), and the orbital angular momenta $(l_{\rho},m_{\rho})$, $(l_{\lambda},m_{\lambda})$, $(l'_{\rho},m'_{\rho})$, and $(l'_{\lambda},m'_{\lambda})$ are defined under the JC-3 in Fig.~\ref{f1}. Using the infinitesimally-shifted Gaussian (ISG) basis
functions, we obtain
\begin{eqnarray}
\notag
 &&[V(\rho_{k})]_{(n_{\rho}n_{\lambda},n_{\rho}'n_{\lambda}')}\\
 \notag
&&=\langle \phi^{G}_{n_{\rho_{3}}l_{\rho_{3}}m_{\rho_{3}}}\phi^{G}_{n_{\lambda_{3}}l_{\lambda_{3}}m_{\lambda_{3}}}|
V(\rho_{k})|\phi^{G}_{n'_{\rho_{3}}l'_{\rho_{3}}m'_{\rho_{3}}}\phi^{G}_{n'_{\lambda_{3}}l'_{\lambda_{3}}m'_{\lambda_{3}}}\rangle\\
 \notag
&&=\{N_{nl}\}\{\lim_{\varepsilon\rightarrow0}\frac{1}{(\nu_{n}\varepsilon)^{l}}\}\sum_{\{\tilde{k}\}}\{C_{lm,\tilde{k}}\}\langle e^{-\nu_{n_{\rho}}(\boldsymbol\rho-\varepsilon_{\rho}\mathbf{D}_{\rho})}e^{-\nu_{n_{\lambda}}(\boldsymbol\lambda-\varepsilon_{\lambda}\mathbf{D}_{\lambda})}|\\
&&V(\rho_{k})|e^{-\nu_{n_{\rho'}}(\boldsymbol\rho-\varepsilon_{\rho'}\mathbf{D}_{\rho'})}
e^{-\nu_{n_{\lambda'}}(\boldsymbol\lambda-\varepsilon_{\lambda'}\mathbf{D}_{\lambda'})}\rangle.
\label{e31}
\end{eqnarray}
Here, $\{\cdot\cdot\cdot\}$ denotes the product of the contained elements. $\sum_{\{\tilde{k}\}}$ means sum over all the $\tilde{k}$ values.

For the final integral of Eq.~(\ref{e31}), the following Jacobi coordinates transformations are performed,
\begin{eqnarray}
\notag
 &&\boldsymbol\rho=\boldsymbol\rho(\boldsymbol\rho_{k},\boldsymbol\lambda_{k})\\
 \notag
&&\boldsymbol\lambda=\boldsymbol\lambda(\boldsymbol\rho_{k},\boldsymbol\lambda_{k})\\
&&d\boldsymbol\rho d\boldsymbol\lambda=\|J\|d\boldsymbol\rho_{k} d\boldsymbol\lambda_{k},
\end{eqnarray}
with $\boldsymbol\rho\equiv\boldsymbol\rho_{3}$, $\boldsymbol\lambda\equiv\boldsymbol\lambda_{3}$, and $k=1,~2,~3$. Here $|J|$ is the Jacobian determinant. The detailed derivation can be found in Ref.~\cite{F602}.

With the help of the ISG basis functions, the matrix elements of the Hamiltonian terms $H_{0}$, $G_{ij}'$, $\tilde{S}_{ij}$, $\tilde{H}^{tensor}_{ij}$, and $\tilde{H}^{c}_{ij}$ can be evaluated directly. The detailed results can be found in Ref.~\cite{F502}.

\subsection{Spin-orbit terms}\label{C}
In Eq.~(\ref{e5}) of Sec.~\ref{sec2.1}, the spin-orbit term $H^{SO(v)}_{ij}$ reads,
\begin{eqnarray}
\notag
\tilde{H}^{so(v)}_{ij} &&=\frac{\textbf{s}_{i}\cdot(\mathbf{r}_{ij}\times\mathbf{p}_{i})}{2m^{2}_{i}r_{ij}}\frac{\partial\tilde{G}^{so(v)}_{ii}}{\partial{r_{ij}}}+
\frac{\textbf{s}_{j}\cdot(-\mathbf{r}_{ij}\times\mathbf{p}_{j})}{2m^{2}_{j}r_{ij}}\frac{\partial\tilde{G}^{so(v)}_{jj}}{\partial{r_{ij}}}+\\
\notag
&&+\frac{[\textbf{s}_{i}\cdot(-\mathbf{r}_{ij}\times\mathbf{p}_{j})+\textbf{s}_{j}\cdot(\mathbf{r}_{ij}\times\mathbf{p}_{i})]}{m_{i}m_{j}r_{ij}}\frac{\partial\tilde{G}^{so(v)}_{ij}}{\partial{r_{ij}}}\\
&&\equiv\tilde{H}^{so(v)ii}_{ij}+\tilde{H}^{so(v)jj}_{ij}+\tilde{H}^{so(v)ij}_{ij}.
\label{e33}
\end{eqnarray}
The Jacobi coordinates transformations are denoted as
\begin{eqnarray}
 \begin{aligned}
 &\mathbf{r}_{ij}=A_{rij}\boldsymbol\rho+B_{rij}\boldsymbol\lambda,\\
 &\mathbf{p}_{i}=A_{pi}\mathbf{p}_{\rho}+B_{pi}\mathbf{p}_{\lambda},
 \end{aligned}
\end{eqnarray}
with $\boldsymbol\rho_{3}\equiv \boldsymbol\rho$ and $\boldsymbol\lambda_{3}\equiv \boldsymbol\lambda$. $A_{rij}$, $B_{rij}$, $A_{pi}$ and $B_{pi}$ can be obtained by Eq.~(\ref{e13}).  Then, the spin-orbit term can be expressed in terms of the Jacobi coordinates $\boldsymbol\rho$ and $\boldsymbol\lambda$, taking the first part of the spin-orbit term as an example,
\begin{eqnarray}
 \begin{aligned}
 &\tilde{H}^{SO(v)ii}_{ij}=\frac{\partial\tilde{G}^{so(v)}_{ii}}{r_{ij}\partial{r_{ij}}}[\frac{A_{rij}A_{pi}}{2m^{2}_{i}}\textbf{\emph{l}}_{\rho}\cdot\mathbf{s}_{i}
 +\frac{B_{rij}B_{pi}}{2m^{2}_{i}}\textbf{\emph{l}}_{\lambda}\cdot\mathbf{s}_{i}\\
  &+\frac{A_{rij}B_{pi}}{2m^{2}_{i}}(\boldsymbol\rho\times\mathbf{p}_{\lambda})\cdot\mathbf{s}_{i}
  +\frac{B_{rij}A_{pi}}{2m^{2}_{i}}(\boldsymbol\lambda\times\mathbf{p}_{\rho})\cdot\mathbf{s}_{i}].
 \end{aligned}
\end{eqnarray}
The terms proportional to $\boldsymbol\lambda\times\mathbf{p}_{\rho}$ or $\boldsymbol\rho\times\mathbf{p}_{\lambda}$ are the three-body spin-orbit potentials, which have no contributions to the current calculations. The reason lies in the following result. According to the Wigner-Eckhart theorem, in the derivation of the matrix elements  $\langle(\tilde{n}\alpha)^{G}_{3}|\frac{\partial\tilde{G}^{so(v)}_{ii}}{r_{ij}\partial{r_{ij}}}(\boldsymbol\rho\times\mathbf{p}_{\lambda})\cdot\mathbf{s}_{i}
 |(\tilde{n}'\alpha)^{G}_{3}\rangle$, a reduced matrix element$\langle l_{\rho}l_{\lambda}L\|\frac{\partial\tilde{G}^{so(v)}_{ii}}{r_{ij}\partial{r_{ij}}}\boldsymbol\rho\times\mathbf{p}_{\lambda}\|l_{\rho}l_{\lambda}L\rangle$ appears and has the
 following form,
\begin{eqnarray}
\notag
 &&\langle l_{\rho}l_{\lambda}L\|\frac{\partial\tilde{G}^{so(v)}_{ii}}{r_{ij}\partial{r_{ij}}}\boldsymbol\rho\times\mathbf{p}_{\lambda}\|l_{\rho}l_{\lambda}L\rangle\\
 \notag
&&=\sqrt{3}(2L+1)X\begin{pmatrix}l_{\rho} & l_{\lambda} & L \\1 & 1 &1 \\l_{\rho} & l_{\lambda} & L \end{pmatrix}\langle l_{\rho}\|\frac{\partial\tilde{G}^{so(v)}_{ii}}{r_{ij}\partial{r_{ij}}}\boldsymbol\rho\|l_{\rho}\rangle\langle l_{\lambda}\|\mathbf{p}_{\lambda}\|l_{\lambda}\rangle,\\
\label{e36}
\end{eqnarray}
where $X(\cdot\cdot\cdot)$ is a 9-j coefficient. $\frac{\partial\tilde{G}^{so(v)}_{ii}}{r_{ij}\partial{r_{ij}}}$, $\boldsymbol\rho$ and $\mathbf{p}_{\lambda}$ are the irreducible spherical tensors of rank  0, 1 and 1, respectively. The 9-j coefficient has an important property, i.e., the result is one factor $(-1)^{\sum{l_{i}}}$ more than the original value, if any two rows (or columns) are permuted. Here $\sum l_{i}$ means sum over all the 9 elements. So, $X(\cdot\cdot\cdot)$ ends up being zero in Eq.~(\ref{e36}).

Hence, the matrix element of $\tilde{H}^{so(v)ii}_{ij}$ in a certain baryon state is expressed,
\begin{eqnarray}
\notag
 &&\langle(\tilde{n},\alpha)^{G}_{3}|\tilde{H}^{SO(v)ii}_{ij}|(\tilde{n}',\alpha)^{G}_{3}\rangle\\
 \notag
&&=\langle(\tilde{n},\alpha)^{G}_{3}|\frac{\partial\tilde{G}^{so(v)}_{ii}}{r_{ij}\partial{r_{ij}}}[\frac{A_{rij}A_{pi}}{2m^{2}_{i}}\textbf{\emph{l}}_{\rho}\cdot\mathbf{s}_{i}
 +\frac{B_{rij}B_{pi}}{2m^{2}_{i}}\textbf{\emph{l}}_{\lambda}\cdot\mathbf{s}_{i}]|(\tilde{n}',\alpha)^{G}_{3}\rangle\\
 \notag
&&\equiv\sum_{\{m_{\xi},m_{\xi}'\}}\{CG_{\xi} CG_{\xi'}\} [(\tilde{H}^{SO(v)ii}_{ij(1)})_{(n_{\rho}n_{\lambda},n_{\rho}'n_{\lambda}');(m_{s_{1,2,3}},m_{s_{1,2,3}}')}\\
&&+(\tilde{H}^{SO(v)ii}_{ij(2)})_{(n_{\rho}n_{\lambda},n_{\rho}'n_{\lambda}');(m_{s_{1,2,3}},m_{s_{1,2,3}}')}],
\end{eqnarray}
with
\begin{eqnarray}
\notag
 &&(\tilde{H}^{SO(v)ii}_{ij(1)})_{(n_{\rho}n_{\lambda},n_{\rho}'n_{\lambda}');(m_{s_{1,2,3}},m_{s_{1,2,3}}')}\\
&&=[\frac{\partial\tilde{G}^{so(v)}_{ii}}{r_{ij}\partial{r_{ij}}}(\frac{A_{rij}A_{pi}}{2m^{2}_{i}}\textbf{\emph{l}}_{\rho}\cdot\mathbf{s}_{i})
]_{(n_{\rho}n_{\lambda},n_{\rho}'n_{\lambda}');(m_{s_{1,2,3}},m_{s_{1,2,3}}')}.
\label{e38}
\end{eqnarray}
The calculation of Eq.~(\ref{e38}) is done in two steps. First, the algebraic calculation of $\textbf{\emph{l}}_{\rho}\cdot\mathbf{s}_{i}$ is performed,
\begin{eqnarray}
\notag
 &&(\textbf{\emph{l}}_{\rho}\cdot\mathbf{s}_{i})|s_{1}m_{s_{1}}'\rangle|s_{2}m_{s_{2}}'\rangle|s_{3}m_{s_{3}}'\rangle|(n_{\rho}'l_{\rho}m_{\rho}')^{G} \rangle|(n_{\lambda}'l_{\lambda}m_{\lambda}')^{G} \rangle\\
 \notag
&&=\sum_{\kappa}\xi_{\kappa}\times(|s_{1}m_{s_{1}}''\rangle|s_{2}m_{s_{2}}''\rangle|s_{3}m_{s_{3}}''\rangle|(n_{\rho}'l_{\rho}m_{\rho}'')^{G} \rangle|(n_{\lambda}'l_{\lambda}m_{\lambda}'')^{G} \rangle)_{\kappa}.\\
&&
\end{eqnarray}
Second, the remaining part with $\frac{\partial\tilde{G}^{so(v)}_{ii}}{r_{ij}\partial{r_{ij}}}$ in Eq.~(\ref{e38}) is finished by means of the ISG basis functions and the Jacobi coordinates transformation $(\boldsymbol\rho_{3}$,$\boldsymbol\lambda_{3})$ $\rightarrow$ $(\boldsymbol\rho_{k}$,$\boldsymbol\lambda_{k})$ ($k$=1,2,3).
In this way, all the matrix elements of the spin-orbit terms can be computed rigorously.

\subsection{Tables of the results}\label{D}
\begin{table*}[htbp]
\begin{ruledtabular}\caption{Contribution of each Hamiltonian term to the mass values (in MeV) for the $1S$-, $1P$- and $1D$-wave states of the $\Lambda_{c}$ and $\Sigma_{c}$ baryons with $\langle H_{mode}\rangle\equiv\langle H_{0}+H^{conf}\rangle$ and $\langle H_{ij}\rangle\equiv\langle H\rangle-\langle(H-H_{ij})\rangle$. The orbital excited states of the $\rho$-mode are marked in bold type. }
\label{tta01}
\begin{tabular}{c c c c c c c c c c c c c c c c c c c}
$(l_{\rho},l_{\lambda})nL(J^{P})_{j}$ & $\langle H_{mode}\rangle$ & $\{\langle H^{t}_{12}\rangle$ & $\langle H^{t}_{23}\rangle$ & $\langle H^{t}_{31}\rangle\}$ & $\{\langle H^{c}_{12}\rangle$  & $\langle H^{c}_{23}\rangle$ & $\langle H^{c}_{31}\rangle\}$ & $\{\langle H^{SO(v)}_{12}\rangle$ & $\langle H^{SO(v)}_{23}\rangle$ & $\langle H^{SO(v)}_{31}\rangle\}$ & $\{\langle H^{SO(s)}_{12}\rangle$ & $\langle H^{SO(s)}_{23}\rangle$& $\langle H^{SO(s)}_{31}\rangle\}$& $\langle H\rangle$  \\\hline
\multicolumn{15}{c}{$\Lambda_{c}$}   \\
$(0,0)1S(\frac{1}{2}^{+})_{0}$ & 2464.30  &$\{$ 0 & 0 & 0 $\}$&$\{$ -176.49  & 0 & 0 $\}$&$\{$ 0 & 0 & 0 $\}$&$\{$ 0 & 0 & 0 $\}$& 2287.81 \\
$(0,1)1P(\frac{1}{2}^{-})_{1}$ & 2781.78  & $\{$ 0 & 0 & 0 $\}$&$\{$ -162.80  & 0 & 0 $\}$&$\{$ 0 & -15.52 & -15.52 $\}$&$\{$ 0 & 3.84 & 3.84 $\}$& 2596.87  \\
$(0,1)1P(\frac{3}{2}^{-})_{1}$ & 2781.78  &$\{$ 0 & 0 & 0 $\}$&$\{$ -161.42  & 0 & 0 $\}$&$\{$ 0 & 7.32 & 7.32 $\}$&$\{$ 0 & -1.86 & -1.86 $\}$& 2630.92  \\
$(0,2)1D(\frac{3}{2}^{+})_{2}$ & 3041.20  &$\{$ 0 & 0 & 0 $\}$&$\{$ -156.64  & 0 & 0 $\}$&$\{$ 0 & -10.51 & -10.51 $\}$&$\{$ 0 & 4.40 & 4.40 $\}$& 2872.53  \\
$(0,2)1D(\frac{5}{2}^{+})_{2}$ & 3041.20  &$\{$ 0 & 0 & 0 $\}$&$\{$ -156.61  & 0 & 0 $\}$&$\{$ 0 & 6.61 & 6.61 $\}$&$\{$ 0 & -2.86 & -2.86 $\}$& 2892.15  \\  \hline
\multicolumn{15}{c}{$\Sigma_{c}$}   \\
$(0,0)1S(\frac{1}{2}^{+})_{1}$ & 2464.30  &$\{$ 0 & 0 & 0 $\}$&$\{$ 48.04  & -27.58 & -27.58 $\}$&$\{$ 0 & 0 & 0 $\}$&$\{$ 0 & 0 & 0 $\}$& 2456.24  \\
$(0,0)1S(\frac{3}{2}^{+})_{1}$ & 2464.30  &$\{$ 0 & 0 & 0 $\}$&$\{$ 44.24  & 11.93 & 11.93 $\}$&$\{$ 0 & 0 & 0 $\}$&$\{$ 0 & 0 & 0 $\}$& 2533.92  \\
$(0,1)1P(\frac{1}{2}^{-})_{0}$ & 2781.78  &$\{$ 0 & 0 & 0 $\}$&$\{$ 42.06  & 0 & 0 $\}$&$\{$ 0 & -43.18 & -43.18 $\}$&$\{$ 0 & 17.15 & 17.15 $\}$& 2773.06  \\
$(0,1)1P(\frac{1}{2}^{-})_{1}$ & 2781.78  &$\{$ 0 & 0 & 0 $\}$&$\{$ 41.80  & -4.72 & -4.72 $\}$&$\{$ 0 & -29.27 & -29.27 $\}$&$\{$ 0 & 10.53 & 10.53 $\}$& 2778.02  \\
$(0,1)1P(\frac{3}{2}^{-})_{1}$ & 2781.78  &$\{$ 0 & 0.74 & 0.74 $\}$&$\{$ 41.13  & 2.14 & 2.14 $\}$&$\{$ 0 & -16.78 & -16.78 $\}$&$\{$ 0 & 7.79 & 7.79 $\}$& 2810.40  \\
$(0,1)1P(\frac{3}{2}^{-})_{2}$ & 2781.78  &$\{$ 0 & -0.44 & -0.44 $\}$&$\{$ 40.65  & -6.02 & -6.02 $\}$&$\{$ 0 & 9.38 & 9.38 $\}$&$\{$ 0 & -6.02 & -6.02 $\}$& 2816.13  \\
$\mathbf{(1,0)}1P(\frac{1}{2}^{-})_{1}$ & 2874.52 &$\{$ 0 & 0 & 0 $\}$&$\{$ -13.81  & 0 & 0 $\}$&$\{$ 0 & -16.64 & -16.64 $\}$&$\{$ 0 & 0 & 0 $\}$& \textbf{2828.13}  \\
$(0,1)1P(\frac{5}{2}^{-})_{2}$ & 2781.78  &$\{$ 0 & 1.38 & 1.38 $\}$&$\{$ 39.79  & 3.38 & 3.38 $\}$&$\{$ 0 & 25.01 & 25.01 $\}$&$\{$ 0 & -10.68 & -10.68 $\}$& 2862.97  \\
$\mathbf{(1,0)}1P(\frac{3}{2}^{-})_{1}$ & 2874.52 &$\{$ 0 & 0 & 0 $\}$&$\{$ -13.11  & 0 & 0 $\}$&$\{$ 0 & 8.07 & 8.07 $\}$&$\{$ 0 & 0 & 0 $\}$& \textbf{2877.37}  \\
$(0,2)1D(\frac{1}{2}^{+})_{1}$ & 3041.20  &$\{$ 0 & 0 & 0 $\}$&$\{$ 39.77  & 1.72 & 1.72 $\}$&$\{$ 0 & -42.03 & -42.03 $\}$&$\{$ 0 & 23.07 & 23.07 $\}$& 3048.14  \\
$(0,2)1D(\frac{3}{2}^{+})_{1}$ & 3041.20  &$\{$ 0 & -0.73 & -0.73 $\}$&$\{$ 39.72  & -0.79 & -0.79 $\}$&$\{$ 0 & -24.37 & -24.37 $\}$&$\{$ 0 & 16.41 & 16.41 $\}$& 3062.98  \\
$(0,2)1D(\frac{3}{2}^{+})_{2}$ & 3041.20  &$\{$ 0 & -0.14 & -0.14 $\}$&$\{$ 39.28  & -0.76 & -0.76 $\}$&$\{$ 0 & -18.78 & -18.78 $\}$&$\{$ 0 & 9.95 & 9.95 $\}$& 3061.57  \\
$(0,2)1D(\frac{5}{2}^{+})_{2}$ & 3041.20  &$\{$ 0 & 0.46 & 0.46 $\}$&$\{$ 39.22  & 0.44 & 0.44 $\}$&$\{$ 0 & -3.66 & -3.66 $\}$&$\{$ 0 & 3.90 & 3.90 $\}$& 3082.51  \\
$(0,2)1D(\frac{5}{2}^{+})_{3}$ & 3041.20  &$\{$ 0 & -0.64 & -0.64 $\}$&$\{$ 38.59  & -1.65 & -1.65 $\}$&$\{$ 0 & 9.43 & 9.43 $\}$&$\{$ 0 & -8.91 & -8.91 $\}$& 3076.68  \\
$(0,2)1D(\frac{7}{2}^{+})_{3}$ & 3041.20  &$\{$ 0 & 1.29 & 1.29 $\}$&$\{$ 38.53  & 1.05 & 1.05 $\}$&$\{$ 0 & 23.17 & 23.17 $\}$&$\{$ 0 & -15.54 & -15.54 $\}$& 3101.93  \\
\end{tabular}
\end{ruledtabular}
\end{table*}

\begin{table*}[htbp]
\begin{ruledtabular}\caption{Calculated $\langle r_{\rho}^{2}\rangle^{1/2}$, $\langle r_{\lambda}^{2}\rangle^{1/2}$ (in fm) and mass values (in MeV) for the $1S$-, $2S$-, $3S$-, $1P$-, $2P$- and $1D$-wave states of the $\Lambda_{c(b)}$ and $\Xi_{c(b)}$ baryons. The experimental data are also listed for comparison, taken by their isospin averages.}
\label{tta1}
\begin{tabular}{c c c c c c c c c c c}
$(l_{\rho},l_{\lambda})nL(J^{P})_{j}$ & $\langle r_{\rho}^{2}\rangle^{1/2}$ & $\langle r_{\lambda}^{2}\rangle^{1/2}$ & $M_{cal.}$ & Baryon/$M_{exp.}/J^{P}_{exp.}$ & $\langle r_{\rho}^{2}\rangle^{1/2}$ & $\langle r_{\lambda}^{2}\rangle^{1/2}$ & $M_{cal.}$ & Baryon/$M_{exp.}/J^{P}_{exp.}$ \\ \hline
& \multicolumn{4}{c}{$\Lambda_{c}$}   &\multicolumn{4}{c}{$\Lambda_{b}$} \\\cline{2-5} \cline{6-9}
$(0,0)1S(\frac{1}{2}^{+})_{0}$ & 0.512  & 0.444 & 2288 & $\Lambda_{c}^{+}$/$\sim$2286/$\frac{1}{2}^{+}$~\cite{F201} & 0.519  & 0.407 & 5622 & $\Lambda_{b}^{0}$/$\sim$5620/$\frac{1}{2}^{+}$~\cite{F201} \\
$(0,0)2S(\frac{1}{2}^{+})_{0}$ & 0.631  & 0.786 & 2764 & $\Lambda_{c}(2765)^{+}$/$\sim$2767/$?^{?}$~\cite{F201} & 0.599  & 0.716 & 6041 & $\Lambda_{b}(6070)^{0}$/$\sim$6072/$\frac{1}{2}^{+}$~\cite{F201} \\
$(0,0)3S(\frac{1}{2}^{+})_{0}$ & 0.988  & 0.633 & 3022 & - & 0.953  & 0.677 & 6352 & -  \\\\
$(0,1)1P(\frac{1}{2}^{-})_{1}$ & 0.541  & 0.633 & 2597 & $\Lambda_{c}(2595)^{+}$/$\sim$2592/$\frac{1}{2}^{-}$~\cite{F201} & 0.536  & 0.579 & 5899 & $\Lambda_{b}(5912)^{0}$/$\sim$5912/$\frac{1}{2}^{-}$~\cite{F201} \\
$(0,1)1P(\frac{3}{2}^{-})_{1}$ & 0.545  & 0.660 & 2631 & $\Lambda_{c}(2625)^{+}$/$\sim$2628/$\frac{3}{2}^{-}$~\cite{F201} & 0.538  & 0.589 & 5913 & $\Lambda_{b}(5920)^{0}$/$\sim$5920/$\frac{3}{2}^{-}$~\cite{F201} \\
$(0,1)2P(\frac{1}{2}^{-})_{1}$ & 0.607  & 0.963 & 2990 & $\Lambda_{c}(2910)^{+}$/$\sim$\textbf{2914}/$?^{?}$~\cite{F201} & 0.579  & 0.855 & 6239 & -\\
$(0,1)2P(\frac{3}{2}^{-})_{1}$ & 0.602  & 0.991 & 3013 & $\Lambda_{c}(2940)^{+}$/$\sim$\textbf{2940}/$\frac{3}{2}^{-}$~\cite{F201} & 0.577  & 0.861 & 6249 & -\\
$(0,2)1D(\frac{3}{2}^{+})_{2}$ & 0.555  & 0.826 & 2873 & $\Lambda_{c}(2860)^{+}$/$\sim$2856/$\frac{3}{2}^{+}$~\cite{F201} & 0.543  & 0.748 & 6135 & $\Lambda_{b}(6146)^{0}$/$\sim$6146/$\frac{3}{2}^{+}$~\cite{F201} \\
$(0,2)1D(\frac{5}{2}^{+})_{2}$ & 0.556  & 0.851 & 2892 & $\Lambda_{c}(2880)^{+}$/$\sim$2882/$\frac{5}{2}^{+}$~\cite{F201} & 0.544  & 0.758 & 6146 & $\Lambda_{b}(6152)^{0}$/$\sim$6153/$\frac{5}{2}^{+}$~\cite{F201}\\\hline
& \multicolumn{4}{c}{$\Xi_{c}$}   &\multicolumn{4}{c}{$\Xi_{b}$} \\\cline{2-5} \cline{6-9}
$(0,0)1S(\frac{1}{2}^{+})_{0}$ & 0.512  & 0.437 & 2479 & $\Xi_{c}^{+,0}$/$\sim$2469/$\frac{1}{2}^{+}$~\cite{F201} & 0.518  & 0.400 & 5806 & $\Xi_{b}^{0,-}$/$\sim$5795/$\frac{1}{2}^{+}$~\cite{F201} \\
$(0,0)2S(\frac{1}{2}^{+})_{0}$ & 0.645  & 0.768 & 2949 & $\Xi_{c}(2970)^{+,0}$/$\sim$2966/$\frac{1}{2}^{+}$~\cite{F201}  & 0.607  & 0.705 & 6224 & $\Xi_{b}(6227)^{0,-}$/$\sim$6227/$?^{?}$~\cite{F201}  \\
$(0,0)3S(\frac{1}{2}^{+})_{0}$ & 0.968  & 0.607 & 3155 & - & 0.990  & 0.549 & 6480 & -  \\\\
$(0,1)1P(\frac{1}{2}^{-})_{1}$ & 0.544  & 0.628 & 2789 & $\Xi_{c}(2790)^{+,0}$/$\sim$2793/$\frac{1}{2}^{-}$~\cite{F201} & 0.540  & 0.573 & 6084 & $\Xi_{b}(6087)^{0}$/$\sim$6087/$\frac{3}{2}^{-}$~\cite{F201} \\
$(0,1)1P(\frac{3}{2}^{-})_{1}$ & 0.549  & 0.654 & 2820 & $\Xi_{c}(2815)^{+,0}$/$\sim$2818/$\frac{3}{2}^{-}$~\cite{F201} & 0.543  & 0.582 & 6097 & $\Xi_{b}(6100)^{0,-}$/$\sim$6097/$\frac{3}{2}^{-}$~\cite{F201} \\
$(0,1)2P(\frac{1}{2}^{-})_{1}$ & 0.616  & 0.950 & 3177 &- & 0.587  & 0.846 & 6422 & -\\
$(0,1)2P(\frac{3}{2}^{-})_{1}$ & 0.612  & 0.977 & 3199 &- & 0.585  & 0.852 & 6431 & -\\
$(0,2)1D(\frac{3}{2}^{+})_{2}$ & 0.563  & 0.822 & 3061 & $\Xi_{c}(3055)^{+}$/$\sim$3056/$\frac{3}{2}^{+}$~\cite{F201}  & 0.552  & 0.742 & 6318 & $\Xi_{b}(6327)^{0}$/$\sim$6327/$?^{?}$~\cite{F201} \\
$(0,2)1D(\frac{5}{2}^{+})_{2}$ & 0.564  & 0.845 & 3078 & $\Xi_{c}(3080)^{+,0}$/$\sim$3079/$\frac{5}{2}^{+}$~\cite{F201} & 0.553  & 0.752 & 6328 & $\Xi_{b}(6333)^{0}$/$\sim$6333/$?^{?}$~\cite{F201}\\
\end{tabular}
\end{ruledtabular}
\end{table*}

\begin{table*}[htbp]
\begin{ruledtabular}\caption{Calculated $\langle r_{\rho}^{2}\rangle^{1/2}$, $\langle r_{\lambda}^{2}\rangle^{1/2}$ (in fm) and mass values (in MeV) for the $1S$-, $2S$-, $3S$-, $1P$- and $1D$-wave states of the $\Sigma_{c}$ and $\Sigma_{b}$ baryons. The orbital excited states of the $\rho$-mode are marked in bold type. The experimental data are also listed for comparison, taken by their isospin averages.}
\label{tta2}
\begin{tabular}{c c c c c c c c c c c}
$(l_{\rho},l_{\lambda})nL(J^{P})_{j}$ & $\langle r_{\rho}^{2}\rangle^{1/2}$ & $\langle r_{\lambda}^{2}\rangle^{1/2}$ & $M_{cal.}$ & Baryon/$M_{exp.}/J^{P}_{exp.}$ & $\langle r_{\rho}^{2}\rangle^{1/2}$ & $\langle r_{\lambda}^{2}\rangle^{1/2}$ & $M_{cal.}$ & Baryon/$M_{exp.}/J^{P}_{exp.}$ \\ \hline
& \multicolumn{4}{c}{$\Sigma_{c}$}   &\multicolumn{4}{c}{$\Sigma_{b}$} \\ \cline{2-5} \cline{6-9}
$(0,0)1S(\frac{1}{2}^{+})_{1}$ & 0.611  & 0.450 & 2456 & $\Sigma_{c}(2455)^{++,+,0}$/$\sim$2453/$\frac{1}{2}^{+}$~\cite{F201} & 0.631  & 0.433 & 5821 & $\Sigma_{b}^{+,-}$/$\sim$5813/$\frac{1}{2}^{+}$~\cite{F201} \\
$(0,0)1S(\frac{3}{2}^{+})_{1}$ & 0.645  & 0.493 & 2534 & $\Sigma_{c}(2520)^{++,+,0}$/$\sim$2518/$\frac{3}{2}^{+}$~\cite{F201} & 0.645  & 0.449 & 5849 & $\Sigma_{b}^{*+,-}$/$\sim$5833/$\frac{3}{2}^{+}$~\cite{F201} \\
$(0,0)2S(\frac{1}{2}^{+})_{1}$ & 0.841  & 0.732 & 2913 & - & 0.774  & 0.716 & 6226 & - \\
$(0,0)2S(\frac{3}{2}^{+})_{1}$ & 0.837  & 0.783 & 2967 & - & 0.770  & 0.734 & 6246 & -\\
$(0,0)3S(\frac{1}{2}^{+})_{1}$ & 0.945  & 0.718 & 3109 & - & 1.019  & 0.607 & 6439 & -  \\
$(0,0)3S(\frac{3}{2}^{+})_{1}$ & 0.992  & 0.696 & 3127 & - & 1.041  & 0.594 & 6446 & -  \\\\
$(0,1)1P(\frac{1}{2}^{-})_{0}$ & 0.658  & 0.640 & 2773 & - & 0.652  & 0.593 & 6087 & - \\
$(0,1)1P(\frac{1}{2}^{-})_{1}$ & 0.662  & 0.647 & 2778 &$\Sigma_{c}(2800)^{++,+,0}$/$\sim$2800/$?^{?}$~\cite{F201} & 0.658  & 0.603 & 6092 & $\Sigma_{b}(6097)^{+,-}$/$\sim$6097/$?^{?}$~\cite{F201}\\
$(0,1)1P(\frac{3}{2}^{-})_{1}$ & 0.670  & 0.672 & 2810 & - & 0.661  & 0.613 & 6105 & - \\
$(0,1)1P(\frac{3}{2}^{-})_{2}$ & 0.678  & 0.688 & 2816 & - & 0.673  & 0.636 & 6113 & - \\
$\mathbf{(1,0)}1P(\frac{1}{2}^{-})_{1}$ & 0.857  & 0.486 & 2828 & $\Sigma_{c}(2846)^{0}$/$\sim$2846/$?^{?}$~\cite{F209} & - &- & - & - \\
$(0,1)1P(\frac{5}{2}^{-})_{2}$ & 0.689  & 0.731 & 2863 &  -  & 0.679  & 0.652 & 6133 & - \\
$\mathbf{(1,0)}1P(\frac{3}{2}^{-})_{1}$ & 0.875  & 0.505 & 2877 & - & - & - & - & -  \\\\
$(0,2)1D(\frac{1}{2}^{+})_{1}$ & 0.683  & 0.817 & 3048 & - & 0.667  & 0.755 & 6330 & - \\
$(0,2)1D(\frac{3}{2}^{+})_{1}$ & 0.684  & 0.834 & 3063 & - & 0.668  & 0.761 & 6337 & - \\
$(0,2)1D(\frac{3}{2}^{+})_{2}$ & 0.690  & 0.846 & 3062 & - & 0.675  & 0.778 & 6334 & -  \\
$(0,2)1D(\frac{5}{2}^{+})_{2}$ & 0.691  & 0.871 & 3083 & - & 0.677  & 0.789 & 6345 & - \\
$(0,2)1D(\frac{5}{2}^{+})_{3}$ & 0.700  & 0.891 & 3076 & - & 0.688  & 0.814 & 6338 & - \\
$(0,2)1D(\frac{7}{2}^{+})_{3}$ & 0.702  & 0.923 & 3102 & - & 0.690  & 0.828 & 6351 & -\\
\end{tabular}
\end{ruledtabular}
\end{table*}

\begin{table*}[htbp]
\begin{ruledtabular}\caption{Same as Table~\ref{tta2}, but for the $\Xi'_{c}$ and $\Xi'_{b}$ baryons.  }
\label{tta3}
\begin{tabular}{c c c c c c c c c }
$(l_{\rho},l_{\lambda})nL(J^{P})_{j}$ & $\langle r_{\rho}^{2}\rangle^{1/2}$ & $\langle r_{\lambda}^{2}\rangle^{1/2}$ & $M_{cal.}$ & Baryon/$M_{exp.}/J^{P}_{exp.}$ & $\langle r_{\rho}^{2}\rangle^{1/2}$ & $\langle r_{\lambda}^{2}\rangle^{1/2}$ & $M_{cal.}$ & Baryon/$M_{exp.}/J^{P}_{exp.}$ \\ \hline
& \multicolumn{4}{c}{$\Xi_{c}'$}   &\multicolumn{4}{c}{$\Xi_{b}'$} \\ \cline{2-5} \cline{6-9}
$(0,0)1S(\frac{1}{2}^{+})_{1}$ & 0.584  & 0.435 & 2589 & $\Xi'^{+,0}_{c}$ /$\sim$2578/$\frac{1}{2}^{+}$~\cite{F201} & 0.602  & 0.414 & 5944 & $\Xi_{b}^{'}(5935)^{-}$/$\sim$5935/$\frac{1}{2}^{+}$~\cite{F201} \\
$(0,0)1S(\frac{3}{2}^{+})_{1}$ & 0.614  & 0.474 & 2660 & $\Xi_{c}(2645)^{+,0}$/$\sim$2645/$\frac{3}{2}^{+}$~\cite{F201}& 0.615  & 0.430 & 5971 & $\Xi_{b}(5955)^{0,-}$/$\sim$5954/$\frac{3}{2}^{+}$~\cite{F201}  \\
$(0,0)2S(\frac{1}{2}^{+})_{1}$ & 0.809  & 0.714 & 3046 & -  & 0.739  & 0.699 & 6351 & -\\
$(0,0)2S(\frac{3}{2}^{+})_{1}$ & 0.804  & 0.762 & 3096 & - & 0.735  & 0.715 & 6369 & - \\
$(0,0)3S(\frac{1}{2}^{+})_{1}$ & 0.925  & 0.685 & 3220 & - & 0.999  & 0.570 & 6543 & - \\
$(0,0)3S(\frac{3}{2}^{+})_{1}$ & 0.967  & 0.668 & 3237 & - & 1.017  & 0.561 & 6551 & - \\\\

$(0,1)1P(\frac{1}{2}^{-})_{0}$ & 0.633  & 0.628 & 2906 & $\Xi_{c}(2882)^{0}$/$\sim$2882/$?^{?}$~\cite{F201} & 0.629  & 0.578 & 6214 & - \\
$(0,1)1P(\frac{1}{2}^{-})_{1}$ & 0.636  & 0.634 & 2912 & - & 0.633  & 0.587 & 6218 & -\\
$(0,1)1P(\frac{3}{2}^{-})_{1}$ & 0.644  & 0.658 & 2941 &$\Xi_{c}(2923)^{+,0}$/$\sim$2923/$?^{?}$~\cite{F201,LHCb25}  & 0.636  & 0.596 & 6230 & - \\
$(0,1)1P(\frac{3}{2}^{-})_{2}$ & 0.649  & 0.670 & 2948 & $\Xi_{c}(2930)^{+,0}$/$\sim$2941/$?^{?}$~\cite{F201} & 0.645  & 0.614 & 6237 & - \\
$\mathbf{(1,0)}1P(\frac{1}{2}^{-})_{1}$ & 0.828  & 0.473 & 2958 & -  & - & - & - & -  \\
$(0,1)1P(\frac{5}{2}^{-})_{2}$ & 0.660  & 0.709 & 2990 & - & 0.650  & 0.629 & 6256 & - \\
$\mathbf{(1,0)}1P(\frac{3}{2}^{-})_{1}$ & 0.847  & 0.490 & 3004 & - & - & - & - & -\\\\
$(0,2)1D(\frac{1}{2}^{+})_{1}$ & 0.660  & 0.808 & 3177 & $\Xi_{c}(3123)^{+}$/$\sim$3123/$?^{?}$~\cite{F201}  & 0.647  & 0.742 & 6452 & -\\
$(0,2)1D(\frac{3}{2}^{+})_{1}$ & 0.662  & 0.824 & 3189 & - & 0.647  & 0.748 & 6458 & -\\
$(0,2)1D(\frac{3}{2}^{+})_{2}$ & 0.666  & 0.833 & 3190 & - & 0.653  & 0.761 & 6456 & -\\
$(0,2)1D(\frac{5}{2}^{+})_{2}$ & 0.668  & 0.856 & 3208 & - & 0.655  & 0.771 & 6466 & -\\
$(0,2)1D(\frac{5}{2}^{+})_{3}$ & 0.674  & 0.870 & 3207 & - & 0.663  & 0.790 & 6461 & - \\
$(0,2)1D(\frac{7}{2}^{+})_{3}$ & 0.676  & 0.899 & 3229 & - & 0.665  & 0.804 & 6473 & - \\
\end{tabular}
\end{ruledtabular}
\end{table*}

\begin{table*}[htbp]
\begin{ruledtabular}\caption{Same as Table~\ref{tta2}, but for the $\Omega_{c}$ and $\Omega_{b}$ baryons. }
\label{tta4}
\begin{tabular}{c c c c c c c c c }
$(l_{\rho},l_{\lambda})nL(J^{P})_{j}$ & $\langle r_{\rho}^{2}\rangle^{1/2}$ & $\langle r_{\lambda}^{2}\rangle^{1/2}$ & $M_{cal.}$ & Baryon/$M_{exp.}/J^{P}_{exp.}$ & $\langle r_{\rho}^{2}\rangle^{1/2}$ & $\langle r_{\lambda}^{2}\rangle^{1/2}$ & $M_{cal.}$ & Baryon/$M_{exp.}/J^{P}_{exp.}$ \\ \hline
& \multicolumn{4}{c}{$\Omega_{c}$}   &\multicolumn{4}{c}{$\Omega_{b}$} \\ \cline{2-5} \cline{6-9}
$(0,0)1S(\frac{1}{2}^{+})_{1}$ & 0.549  & 0.417 & 2696 & $\Omega_{c}^{0}$/$\sim$2695/$\frac{1}{2}^{+}$~\cite{F201} & 0.564  & 0.395 & 6043 & $\Omega_{b}^{-}$/$\sim$6045/$\frac{1}{2}^{+}$~\cite{F201}  \\
$(0,0)1S(\frac{3}{2}^{+})_{1}$ & 0.578  & 0.454 & 2765 & $\Omega_{c}(2770)^{0}$/$\sim$2766/$\frac{3}{2}^{+}$~\cite{F201}  & 0.576  & 0.409 & 6069 & - \\
$(0,0)2S(\frac{1}{2}^{+})_{1}$ & 0.775  & 0.686 & 3150 & - & 0.705  & 0.672 & 6448 & -\\
$(0,0)2S(\frac{3}{2}^{+})_{1}$ & 0.771  & 0.730 & 3198 & $\Omega_{c}(3185)^{0}$/$\sim$3185/$?^{?}$~\cite{F201} & 0.702  & 0.687 & 6465 & - \\
$(0,0)3S(\frac{1}{2}^{+})_{1}$ & 0.882  & 0.672 & 3325 & $\Omega_{c}(3327)^{0}$/$\sim$3327/$?^{?}$~\cite{F201} & 0.953  & 0.560 & 6641 & -\\
$(0,0)3S(\frac{3}{2}^{+})_{1}$ & 0.924  & 0.654 & 3339 & - & 0.973  & 0.549 & 6647 & -\\\\
$(0,1)1P(\frac{1}{2}^{-})_{0}$ & 0.602  & 0.605 & 3009 &$\Omega_{c}(3000)^{0}$/$\sim$3000/$?^{?}$~\cite{F201}  & 0.595  & 0.552 & 6308 & $\Omega_{b}(6316)^{-}$/$\sim$6315/$?^{?}$~\cite{F201} \\
$(0,1)1P(\frac{1}{2}^{-})_{1}$ & 0.604  & 0.609 & 3015 & - & 0.599  & 0.560 & 6313 & -\\
$(0,1)1P(\frac{3}{2}^{-})_{1}$ & 0.612  & 0.633 & 3045 & $\Omega_{c}(3050)^{0}$/$\sim$3050/$?^{?}$~\cite{F201}   & 0.602  & 0.570 & 6326 & $\Omega_{b}(6330)^{-}$/$\sim$6330/$?^{?}$~\cite{F201}\\
$(0,1)1P(\frac{3}{2}^{-})_{2}$ & 0.615  & 0.643 & 3052 & - & 0.608  & 0.586 & 6334 & $\Omega_{b}(6340)^{-}$/$\sim$6340/$?^{?}$~\cite{F201}\\
$\mathbf{(1,0)}1P(\frac{1}{2}^{-})_{1}$ & 0.792  & 0.459 & 3059 & $\Omega_{c}(3065)^{0}$/$\sim$3065/$?^{?}$~\cite{F201} & - & - & -& - \\
$(0,1)1P(\frac{5}{2}^{-})_{2}$ & 0.626  & 0.683 & 3095 & $\Omega_{c}(3090)^{0}$/$\sim$3090/$?^{?}$~\cite{F201} & 0.614  & 0.601 & 6353 & $\Omega_{b}(6350)^{-}$/$\sim$6350/$?^{?}$~\cite{F201} \\
$\mathbf{(1,0)}1P(\frac{3}{2}^{-})_{1}$ & 0.813  & 0.479 & 3109 &$\Omega_{c}(3120)^{0}$/$\sim$3119/$?^{?}$~\cite{F201} & - & - & - & - \\\\
$(0,2)1D(\frac{1}{2}^{+})_{1}$ & 0.631  & 0.782 & 3278 & - & 0.616  & 0.713 & 6544 & - \\
$(0,2)1D(\frac{3}{2}^{+})_{1}$ & 0.633  & 0.801 & 3292 & - & 0.617  & 0.720 & 6552 & -\\
$(0,2)1D(\frac{3}{2}^{+})_{2}$ & 0.635  & 0.806 & 3293 & - & 0.621  & 0.731 & 6550 & -\\
$(0,2)1D(\frac{5}{2}^{+})_{2}$ & 0.637  & 0.831 & 3311 & - & 0.622  & 0.742 & 6561 & -\\
$(0,2)1D(\frac{5}{2}^{+})_{3}$ & 0.640  & 0.840 & 3310 & - & 0.627  & 0.759 & 6557 & -\\
$(0,2)1D(\frac{7}{2}^{+})_{3}$ & 0.642  & 0.871 & 3332 & - & 0.629  & 0.772 & 6570 & -\\
\end{tabular}
\end{ruledtabular}
\end{table*}

\begin{table*}[htbp]
\begin{ruledtabular}\caption{The deviations of the calculated masses of the 74 baryons from the measured ones~\cite{F201,F209,LHCb25}. Most of the deviations are less than 20 MeV. The arithmetic average deviation $(\sum_{i=1}^{n}|M_{cal.}-M_{exp.}|_{i})/n$ is about 9.12 MeV. $M_{exp.}$ denotes the central value of the measured mass. `$\uparrow$' means the same as above. The $\Lambda_{c}(2910)^{+}$, $\Lambda_{c}(2940)^{+}$ and $\Xi_{c}(3123)^{+}$ are not included in the list. }
\label{a8}
\begin{tabular}{c c c c c | c c c c c }
Baryon ($J^{P}$) & $M_{exp.}$  & $(l_{\rho},l_{\lambda})nL(J^{P})_{j}$ & $M_{cal.}$ & $M_{cal.}$-$M_{exp.}$  & Baryon ($J^{P}$) & $M_{exp.}$  & $(l_{\rho},l_{\lambda})nL(J^{P})_{j}$ & $M_{cal.}$ & $M_{cal.}$-$M_{exp.}$ \\ \hline
$\Lambda_{c}^{+}(\frac{1}{2}^{+})$ & 2286.46  & $(0,0)1S(\frac{1}{2}^{+})_{0}$ & 2288 & 1.54 & $\Omega_{c}(2770)^{0}(\frac{3}{2}^{+})$ & 2766  & $(0,0)1S(\frac{3}{2}^{+})_{1}$ & 2765 & -1 \\
$\Lambda_{c}(2595)^{+}(\frac{1}{2}^{-})$ & 2592.25 & $(0,1)1P(\frac{1}{2}^{-})_{1}$ & 2597 & 4.75& $\Omega_{c}(3000)^{0}(?^{?})$ & 3000.46  & $(0,1)1P(\frac{1}{2}^{-})_{0}$ & 3009 & 8.54  \\
$\Lambda_{c}(2625)^{+}(\frac{3}{2}^{-})$ & 2628  & $(0,1)1P(\frac{3}{2}^{-})_{1}$ & 2631 & 3 & $\Omega_{c}(3050)^{0}(?^{?})$ & 3050.17  & $(0,1)1P(\frac{3}{2}^{-})_{1}$ & 3045 & -5.17  \\
$\Lambda_{c}(2765)^{+}(?^{?})$ & 2766.6  & $(0,0)2S(\frac{1}{2}^{+})_{0}$ & 2764 & -2.6 & $\Omega_{c}(3065)^{0}(?^{?})$ & 3065.58  & $(1,0)1P(\frac{1}{2}^{-})_{1}$ & 3059 & -6.58  \\
$\Lambda_{c}(2860)^{+}(\frac{3}{2}^{+})$ & 2856.1  & $(0,2)1D(\frac{3}{2}^{+})_{2}$ & 2873 & 16.9 & $\Omega_{c}(3090)^{0}(?^{?})$ & 3090.15  & $(0,1)1P(\frac{5}{2}^{-})_{2}$ & 3095 & 4.85  \\
$\Lambda_{c}(2880)^{+}(\frac{5}{2}^{+})$ & 2881.62  & $(0,2)1D(\frac{5}{2}^{+})_{2}$ & 2892 & 10.38 & $\Omega_{c}(3120)^{0}(?^{?})$ & 3118.98  & $(1,0)1P(\frac{3}{2}^{-})_{1}$ & 3109 & -9.98  \\
$\Sigma_{c}(2455)^{++}(\frac{1}{2}^{+})$ & 2453.97  & $(0,0)1S(\frac{1}{2}^{+})_{1}$ & 2456 & 2.03 & $\Omega_{c}(3185)^{0}(?^{?})$ & 3185  & $(0,0)2S(\frac{3}{2}^{+})_{1}$ & 3198 & 13  \\
$\Sigma_{c}(2455)^{+}(\frac{1}{2}^{+})$ & 2452.65  & $\uparrow$ & $\uparrow$ & 3.35 & $\Omega_{c}(3327)^{0}(?^{?})$ & 3327.1  & $(0,0)3S(\frac{1}{2}^{+})_{1}$ & 3325 & -2.1  \\
$\Sigma_{c}(2455)^{0}(\frac{1}{2}^{+})$ & 2453.75  & $\uparrow$ & $\uparrow$ & 2.25 & $\Lambda_{b}^{0}(\frac{1}{2}^{+})$ & 5619.5  & $(0,0)1S(\frac{1}{2}^{+})_{0}$ & 5622 & 2.43 \\
$\Sigma_{c}(2520)^{++}(\frac{3}{2}^{+})$ & 2518.42  & $(0,0)1S(\frac{3}{2}^{+})_{1}$ & 2534 & 15.59 & $\Lambda_{b}(5912)^{0}(\frac{1}{2}^{-})$ & 5912.16  & $(0,1)1P(\frac{1}{2}^{-})_{1}$ & 5899 & -13.2 \\
$\Sigma_{c}(2520)^{+}(\frac{3}{2}^{+})$ & 2517.4  & $\uparrow$ & $\uparrow$ & 16.6 & $\Lambda_{b}(5920)^{0}(\frac{3}{2}^{-})$ & 5920.07  & $(0,1)1P(\frac{3}{2}^{-})_{1}$ & 5913 & -7.07 \\
$\Sigma_{c}(2520)^{0}(\frac{3}{2}^{+})$ & 2518.48  & $\uparrow$ & $\uparrow$ & 15.52 & $\Lambda_{b}(6070)^{0}(\frac{1}{2}^{+})$ & 6072.3  & $(0,0)2S(\frac{1}{2}^{+})_{0}$ & 6041 & -31.3 \\
$\Sigma_{c}(2800)^{++}(?^{?})$ & 2801  & $(0,1)1P(\frac{1}{2}^{-})_{1}$ & 2778 & -23& $\Lambda_{b}(6146)^{0}(\frac{3}{2}^{+})$ & 6146.2  & $(0,2)1D(\frac{3}{2}^{+})_{2}$ & 6135 & -11.2 \\
$\Sigma_{c}(2800)^{+}(?^{?})$ & 2792  & $\uparrow$ & $\uparrow$ & -14 & $\Lambda_{b}(6152)^{0}(\frac{5}{2}^{+})$ & 6152.5  & $(0,2)1D(\frac{5}{2}^{+})_{2}$ & 6146 & -6.5 \\
$\Sigma_{c}(2800)^{0}(?^{?})$ & 2806  & $\uparrow$ & $\uparrow$ & -28 &$\Sigma_{b}^{+}(\frac{1}{2}^{+})$   & 5810.56  & $(0,0)1S(\frac{1}{2}^{+})_{1}$ & 5821 & 10.44 \\
$\Sigma_{c}(2846)^{0}(?^{?})$ & 2846  & $(1,0)1P(\frac{1}{2}^{-})_{1}$ & 2828 & -18 & $\Sigma_{b}^{-}(\frac{1}{2}^{+})$ & 5815.64 & $\uparrow$ & $\uparrow$ & 5.36 \\
$\Xi_{c}^{+}(\frac{1}{2}^{+})$ & 2467.95  & $(0,0)1S(\frac{1}{2}^{+})_{0}(\mathbf{\bar{3}}_{F})$ & 2479 & 11.05 & $\Sigma_{b}^{*+}(\frac{3}{2}^{+})$ & 5830.32 & $(0,0)1S(\frac{3}{2}^{+})_{1}$ & 5849 & 18.68 \\
$\Xi_{c}^{0}(\frac{1}{2}^{+})$ & 2470.44  & $\uparrow$ & $\uparrow$ & 8.56 & $\Sigma_{b}^{*-}(\frac{3}{2}^{+})$ & 5834.74 & $\uparrow$ & $\uparrow$ & 14.26 \\
$\Xi_{c}^{'+}(\frac{1}{2}^{+})$ & 2578.2  & $(0,0)1S(\frac{1}{2}^{+})_{1}(\mathbf{6}_{F})$ & 2589 & 10.8 & $\Sigma_{b}(6097)^{+}(?^{?})$ & 6095.8 & $(0,1)1P(\frac{1}{2}^{-})_{1}$ & 6092 & -3.8 \\
$\Xi_{c}^{'0}(\frac{1}{2}^{+})$ & 2578.7  & $\uparrow$ & $\uparrow$ & 10.3 & $\Sigma_{b}(6097)^{-}(?^{?})$ & 6098.0 & $\uparrow$ & $\uparrow$ & -6.0 \\
$\Xi_{c}(2645)^{+}(\frac{3}{2}^{+})$ & 2645.1  & $(0,0)1S(\frac{3}{2}^{+})_{1}(\mathbf{6}_{F})$ & 2660 & 14.9 & $\Xi_{b}^{-}(\frac{1}{2}^{+})$  & 5797 & $(0,0)1S(\frac{1}{2}^{+})_{0}(\mathbf{6}_{F})$ & 5806 & 9 \\
$\Xi_{c}(2645)^{0}(\frac{3}{2}^{+})$ & 2645.7  & $\uparrow$ & $\uparrow$ & 14.3 & $\Xi_{b}^{0}(\frac{1}{2}^{+})$  & 5791.7 & $\uparrow$ & $\uparrow$ & 14.3 \\
$\Xi_{c}(2790)^{+}(\frac{1}{2}^{-})$ & 2791.9  & $(0,1)1P(\frac{1}{2}^{-})_{1}(\mathbf{\bar{3}}_{F})$ & 2789 & -2.9 & $\Xi_{b}(5935)^{-}(\frac{1}{2}^{+})$  & 5934.9 & $(0,0)1S(\frac{1}{2}^{+})_{1}(\mathbf{6}_{F})$ & 5944 & 9.1 \\
$\Xi_{c}(2790)^{0}(\frac{1}{2}^{-})$ & 2793.9  & $\uparrow$ & $\uparrow$ & -4.9 & $\Xi_{b}(5945)^{0}(\frac{3}{2}^{+})$  & 5952.3 & $(0,0)1S(\frac{3}{2}^{+})_{1}(\mathbf{6}_{F})$ & 5971 & 18.7 \\
$\Xi_{c}(2815)^{+}(\frac{3}{2}^{-})$ & 2816.51  & $(0,1)1P(\frac{3}{2}^{-})_{1}(\mathbf{\bar{3}}_{F})$ & 2820 & 3.49 & $\Xi_{b}(5955)^{-}(\frac{3}{2}^{+})$  & 5955.5 & $\uparrow$ & $\uparrow$ & 15.5 \\
$\Xi_{c}(2815)^{0}(\frac{3}{2}^{-})$ & 2819.79  & $\uparrow$ & $\uparrow$ & 0.21 & $\Xi_{b}(6087)^{-}(\frac{3}{2}^{-})$  & 6087 & $(0,1)1P(\frac{1}{2}^{-})_{1}(\mathbf{\bar{3}}_{F})$ & 6084 & -3 \\
$\Xi_{c}(2882)^{0}(?^{?})$ & 2882  & $(0,1)1P(\frac{1}{2}^{-})_{0}(\mathbf{6}_{F})$ & 2906 & 24 & $\Xi_{b}(6095)^{0}(\frac{3}{2}^{-})$  & 6095.1 & $(0,1)1P(\frac{3}{2}^{-})_{1}(\mathbf{\bar{3}}_{F})$ & 6097 & 1.9 \\
$\Xi_{c}(2923)^{+}(?^{?})$ & 2922.8  & $(0,1)1P(\frac{3}{2}^{-})_{1}(\mathbf{6}_{F})$ & 2941 & 18.2 & $\Xi_{b}(6100)^{-}(\frac{3}{2}^{-})$  & 6099.8 & $\uparrow$ & $\uparrow$ & -2.8 \\
$\Xi_{c}(2923)^{0}(?^{?})$ & 2923.2  & $\uparrow$ & $\uparrow$ & 17.8 & $\Xi_{b}(6227)^{-}(?^{?})$  & 6227.9 & $(0,0)2S(\frac{1}{2}^{+})_{0}(\mathbf{\bar{3}}_{F})$ & 6224 & -3.9 \\
$\Xi_{c}(2930)^{+}(?^{?})$ & 2942  & $(0,1)1P(\frac{3}{2}^{-})_{2}(\mathbf{6}_{F})$ & 2948 & 6 & $\Xi_{b}(6227)^{0}(?^{?})$  & 6226.8 & $\uparrow$ & $\uparrow$ & -2.8 \\
$\Xi_{c}(2930)^{0}(?^{?})$ & 2938.55  & $\uparrow$ & $\uparrow$ & 9.45 & $\Xi_{b}(6327)^{0}(?^{?})$ & 6327.28 & $(0,2)1D(\frac{3}{2}^{-})_{2}(\mathbf{\bar{3}}_{F})$ & 6318 & -9.28 \\
$\Xi_{c}(2970)^{+}(\frac{1}{2}^{+})$ & 2964.3  & $(0,0)2S(\frac{1}{2}^{+})_{0}(\mathbf{\bar{3}}_{F})$ & 2949 & -15.3 & $\Xi_{b}(6333)^{0}(?^{?})$  & 6332.69 & $(0,2)1D(\frac{5}{2}^{-})_{2}(\mathbf{\bar{3}}_{F})$ & 6328 & -4.69 \\
$\Xi_{c}(2970)^{0}(\frac{1}{2}^{+})$ & 2965.9  & $\uparrow$ & $\uparrow$ & -16.9 & $\Omega_{b}^{-}(\frac{1}{2}^{+})$  & 6045.8 & $(0,0)1S(\frac{1}{2}^{+})_{1}$ & 6043 & -2.8 \\
$\Xi_{c}(3055)^{+}(\frac{3}{2}^{+})$ & 3055.9  & $(0,2)1D(\frac{3}{2}^{+})_{2}(\mathbf{\bar{3}}_{F})$ & 3061 & 5.1 & $\Omega_{b}(6316)^{-}(?^{?})$  & 6315.6 & $(0,1)1P(\frac{1}{2}^{-})_{0}$ & 6308 & -7.6 \\
$\Xi_{c}(3080)^{+}(\frac{5}{2}^{+})$ & 3077.2  & $(0,2)1D(\frac{5}{2}^{+})_{2}(\mathbf{\bar{3}}_{F})$ & 3078 & 0.8 & $\Omega_{b}(6330)^{-}(?^{?})$ & 6330.3 & $(0,1)1P(\frac{3}{2}^{-})_{1}$ & 6326 & -4.3 \\
$\Xi_{c}(3080)^{0}(\frac{5}{2}^{+})$ & 3079.9  & $\uparrow$ & $\uparrow$ & -1.9 & $\Omega_{b}(6340)^{-}(?^{?})$  & 6339.7 & $(0,1)1P(\frac{3}{2}^{-})_{2}$ & 6334 & -5.7 \\
$\Omega_{c}^{0}(\frac{1}{2}^{+})$ & 2695.3  & $(0,0)1S(\frac{1}{2}^{+})_{1}$ & 2696 & 0.7 & $\Omega_{b}(6350)^{-}(?^{?})$  & 6349.8 & $(0,1)1P(\frac{5}{2}^{-})_{2}$ & 6353 & 3.2 \\
\end{tabular}
\end{ruledtabular}
\end{table*}

\end{document}